\documentclass[iop,revtex4]{emulateapj}
\usepackage{natbib, color}

\begin{document}

\title{Complex Spiral Structure in the HD 100546 Transitional Disk as Revealed by GPI and MagAO}

\author{Katherine B. Follette\altaffilmark{1,2,3}, Julien Rameau\altaffilmark{4}, Ruobing Dong\altaffilmark{5}, Laurent Pueyo\altaffilmark{6}, Laird M. Close\altaffilmark{5}, Gaspard Duch{\^e}ne\altaffilmark{7,8}, Jeffrey Fung\altaffilmark{7,3}, Clare Leonard\altaffilmark{2}, Bruce Macintosh\altaffilmark{1}, Jared R. Males\altaffilmark{5}, Christian Marois\altaffilmark{9,10}, Maxwell A. Millar-Blanchaer\altaffilmark{11,12}, Katie M. Morzinski\altaffilmark{5}, Wyatt Mullen\altaffilmark{1}, Marshall Perrin\altaffilmark{6}, Elijah Spiro\altaffilmark{2}, Jason Wang\altaffilmark{7}, S. Mark Ammons\altaffilmark{13}, Vanessa P. Bailey\altaffilmark{1}, Travis Barman\altaffilmark{14}, Joanna Bulger\altaffilmark{15}, Jeffrey Chilcote\altaffilmark{16}, Tara Cotten\altaffilmark{17}, Robert J. De Rosa\altaffilmark{7}, Rene Doyon\altaffilmark{4}, Michael P. Fitzgerald\altaffilmark{18}, Stephen J. Goodsell\altaffilmark{19,20}, James R. Graham\altaffilmark{7}, Alexandra Z. Greenbaum\altaffilmark{21}, Pascale Hibon\altaffilmark{22}, Li-Wei Hung\altaffilmark{18}, Patrick Ingraham\altaffilmark{23}, Paul Kalas\altaffilmark{7,24}, Quinn Konopacky\altaffilmark{25}, James E. Larkin\altaffilmark{18}, J{\'e}r{\^o}me Maire\altaffilmark{25}, Franck Marchis\altaffilmark{24}, Stanimir Metchev\altaffilmark{26}, Eric L. Nielsen\altaffilmark{1,24}, Rebecca Oppenheimer\altaffilmark{27}, David Palmer\altaffilmark{13}, Jennifer Patience\altaffilmark{28}, Lisa Poyneer\altaffilmark{13}, Abhijith Rajan\altaffilmark{28}, Fredrik T. Rantakyr{\"o}\altaffilmark{29}, Dmitry Savransky\altaffilmark{30}, Adam C. Schneider\altaffilmark{28}, Anand Sivaramakrishnan\altaffilmark{6}, Inseok Song\altaffilmark{17}, Remi Soummer\altaffilmark{6}, Sandrine Thomas\altaffilmark{23}, David Vega\altaffilmark{24}, J. Kent Wallace\altaffilmark{11}, Kimberly Ward-Duong\altaffilmark{28}, Sloane Wiktorowicz\altaffilmark{31}, and Schuyler Wolff\altaffilmark{21}}

\altaffiltext{1}{Kavli Institute for Particle Astrophysics and Cosmology, Department of Physics, Stanford University, Stanford, CA, 94305, USA}
\altaffiltext{2}{Physics and Astronomy Department, Amherst College, 21 Merrill Science Drive, Amherst, MA 01002, USA}
\altaffiltext{3}{NASA Sagan Fellow}
\altaffiltext{4}{Institut de Recherche sur les Exoplan{\`e}tes, D{\'e}partment de Physique, Universit{\'e} de Montr{\'e}al, Montr{\'e}al QC H3C 3J7, Canada}
\altaffiltext{5}{Steward Observatory, University of Arizona, Tucson AZ 85721, USA}
\altaffiltext{6}{Space Telescope Science Institute, Baltimore, MD 21218, USA}
\altaffiltext{7}{Astronomy Department, University of California, Berkeley; Berkeley CA 94720, USA}
\altaffiltext{8}{Univ. Grenoble Alpes/CNRS, IPAG, F-38000 Grenoble, France}
\altaffiltext{9}{National Research Council of Canada Herzberg, 5071 West Saanich Rd, Victoria, BC V9E 2E7, Canada}
\altaffiltext{10}{University of Victoria, 3800 Finnerty Rd, Victoria, BC V8P 5C2, Canada}
\altaffiltext{11}{Jet Propulsion Laboratory, California Institute of Technology Pasadena CA 91125, USA}
\altaffiltext{12}{Hubble Fellow}
\altaffiltext{13}{Lawrence Livermore National Laboratory, Livermore, CA 94551, USA}
\altaffiltext{14}{Lunar and Planetary Laboratory, University of Arizona, Tucson AZ 85721, USA}
\altaffiltext{15}{Subaru Telescope, NAOJ, 650 North A'ohoku Place, Hilo, HI 96720, USA}
\altaffiltext{16}{Dunlap Institute for Astronomy \& Astrophysics, University of Toronto, Toronto, ON M5S 3H4, Canada}
\altaffiltext{17}{Department of Physics and Astronomy, University of Georgia, Athens, GA 30602, USA}
\altaffiltext{18}{Department of Physics \& Astronomy, University of California, Los Angeles, CA 90095, USA}
\altaffiltext{19}{Gemini Observatory, 670 N. A'ohoku Place, Hilo, HI 96720, USA}
\altaffiltext{20}{Department of Physics, Durham University, Stockton Road, Durham DH1, UK}
\altaffiltext{21}{Department of Physics and Astronomy, Johns Hopkins University, Baltimore, MD 21218, USA}
\altaffiltext{22}{European Southern Observatory, Alonso de Cordova 3107, Vitacura, Santiago, Chile}
\altaffiltext{23}{Large Synoptic Survey Telescope, 950N Cherry Av, Tucson, AZ 85719, USA}
\altaffiltext{24}{SETI Institute, Carl Sagan Center, 189 Bernardo Avenue,  Mountain View, CA 94043, USA}
\altaffiltext{25}{Center for Astrophysics and Space Science, University of California San Diego, La Jolla, CA 92093, USA}
\altaffiltext{26}{Department of Physics and Astronomy, Centre for Planetary Science and Exploration, the University of Western Ontario, London, ON N6A 3K7, Canada}
\altaffiltext{27}{American Museum of Natural History, Depratment of Astrophysics, New York, NY 10024, USA}
\altaffiltext{28}{School of Earth and Space Exploration, Arizona State University, PO Box 871404, Tempe, AZ 85287, USA}
\altaffiltext{29}{Gemini Observatory, Casilla 603, La Serena, Chile}
\altaffiltext{30}{Sibley School of Mechanical and Aerospace Engineering, Cornell University, Ithaca, NY 14853, USA}
\altaffiltext{31}{The Aerospace Corporation, 2310 E. El Segundo Blvd., El Segundo, CA 90245}

\begin{abstract}
We present optical and near-infrared high contrast images of the transitional disk HD 100546 taken with the Magellan Adaptive Optics system (MagAO) and the Gemini Planet Imager (GPI). GPI data include both polarized intensity and total intensity imagery, and MagAO data are taken in Simultaneous Differential Imaging mode at H$\alpha$. The new GPI \textit{H}-band total intensity data represent a significant enhancement in sensitivity and field rotation compared to previous data sets and enable a detailed exploration of substructure in the disk. The data are processed with a variety of differential imaging techniques (polarized, angular, reference, and simultaneous differential imaging) in an attempt to identify the disk structures that are most consistent across wavelengths, processing techniques, and algorithmic parameters. The inner disk cavity at 15 au is clearly resolved in multiple datasets, as are a variety of spiral features. While the cavity and spiral structures are identified at levels significantly distinct from the neighboring regions of the disk under several algorithms and with a range of algorithmic parameters, emission at the location of HD 100546 c varies from point-like under aggressive algorithmic parameters to a smooth continuous structure with conservative parameters, and is consistent with disk emission. Features identified in the HD100546 disk bear qualitative similarity to computational models of a moderately inclined two-armed spiral disk, where projection effects and wrapping of the spiral arms around the star result in a number of truncated spiral features in forward-modeled images.    
\end{abstract}


\section{Introduction} 
\label{sec:intro}

Transitional disks were first identified as a circumstellar disk subclass based purely on the peculiar lack of near-infrared (NIR) excess in their spectral energy distributions (SEDs; \citealp{strom89}) relative to full protoplanetary disks. This NIR deficit was hypothesized to result from dust depletion in the inner disk at scales of a few to a few tens of au, and to be an indication that these disks were in the process of transitioning (through disk clearing) to more evolved debris disks, hence their name. Development of large millimeter interferometers and high-resolution NIR Adaptive Optics (AO) systems have since enabled resolved images of centrally-cleared regions in transitional disks at both millimeter and NIR wavelengths. Evidence of ubiquitous disk asymmetries \citep[e.g.,][]{vandermarel13,Follette:2015} and recently confirmation of embedded accreting objects in these disks \citep{close14, Sallum:2015ej} have lent significant fodder to the hypothesis that transitional disk cavities are a result of ongoing planet formation  \citep{Owen:2016}, at least in some cases. 

The disk around the Herbig Ae star HD 100546 (B9Vne, $109\pm 4$ pc, $5$--$10$ Myr \citealt{vanderancker97,Guimaraes06,levenhagen06,vanLeeuwen:2007dc,lindegren:2016}) was first identified through the large infrared excess and prominent crystalline features in the SED \citep{Hu:89, Waelkens:96}. The first resolved images of the HD 100546 disk were obtained in NIR scattered light with an early AO system by \citet{Pantin:00}. They revealed a smooth, bright, elliptical disk extending to $\sim$230 au. Subsequent imaging with the Hubble Space Telescope's NICMOS \citep{Augereau:2001}, STIS \citep{Grady:2001} and ACS \citep{ardilla07} cameras revealed the disk at increasingly high resolution, and showed for the first time distinct disk asymmetries. Due to its bright central star and complex morphology, HD 100546 has been studied extensively, and is the subject of several hundred scientific studies. Therefore, only the most immediately relevant findings to the observations described in this paper are summarized here. We note that the distance to HD100546 was recently measured by GAIA to be 109$\pm$4pc \citep{lindegren:2016}, which is somewhat larger than the previous estimate of 97$\pm$4pc \citep{vanLeeuwen:2007dc}. We have updated numbers in this paper, including those from past literature, to reflect this new distance. 

The HD 100546 disk exhibits complex morphology on a variety of spatial scales. Its features include: a large-scale brightness anisotropy along the disk minor axis \citep{Augereau:2001}, flaring \citep{grady05}, a resolved cavity \citep{avenhaus14, Garufi:2016}, and  prominent spiral arms \citep{boccaletti13,avenhaus14,Currie:2015,Garufi:2016}. The moderate disk inclination (42$^{\circ}$) further complicates the appearance of the disk, with most features being detected to the North and East of the central star, on the illuminated half (back-scattering) of the disk. Due to its inclination, it is likely that the lack of detected near-side disk features in HD 100546 is a result of a scattering phase function with a relatively low forward-scattering efficiency, though projection effects and obscuration by the disk midplane likely also play a role.

The inner disk cavity has been resolved several times with the VLT Interferometer in the NIR, and extends from $0.8$--$15$ au in radius \citep{benisty10,Tatulli:2011,Panic:2014}. The outer edge of this inner disk cavity has since been confirmed by ground based AO Polarimetric Differential Imaging (PDI, e.g., \citealt{Kuhn:01}) in the NIR and visible \citep{quanz11,avenhaus14,Garufi:2016} with estimated cavity radii ranging from 12.5--17au. The most recent, highest-resolution measurements, taken with SPHERE by \citet{Garufi:2016}, suggest that the peak of the inner disk rim may lie slightly farther inward at shorter wavelength (12.5au at \textit{R} versus 15au at \textit{H} and \textit{K}). 

A number of studies have uncovered asymmetric structures in the disk beyond the 15 au inner cavity rim. These include spiral arm-like asymmetries, but these features are stationary over five to nine year periods, inconsistent with launching by a fast-orbiting inner planet candidate \citep{avenhaus14,boccaletti13,Garufi:2016}. Other identified asymmetric disk features include: a small scale spiral arm to the East \citep{Garufi:2016}, and an arc-like feature (``wing") along the disk minor axis \citep{Garufi:2016}. The nature of these structures is not yet well understood. 

While visible and NIR observations probe structures in the disk's surface layers at high resolution, the large particles that make up the disk midplane can only be studied at longer millimeter wavelengths. Millimeter images of the HD 100546 midplane are best reproduced with a two component model: an outer ring centered at $215$ au with a radial extent of $85$ au and an inner, incomplete ring (horseshoe) from $30$ to $60$ au \citep{pineda14,Walsh:14,Wright:15}. The inner rim of the thermally-emitting millimeter dust cavity is thus a factor of $2$--$3$ more distant than the rim of the NIR scattered light cavity.  Observed variations in cavity radius with wavelength have some precedent in transition disks \citep{Dong:2012,Follette:2013,Pinilla:2015b} and can be explained by pressure traps in which large particles are caught while the smallest particles can diffuse closer-in \citep{pinilla12}. This ``dust filtration" phenomenon is also predicted from planet-disk interaction models for relatively low mass planets \citep{Zhu:2012}. 

The disparity between cavity radii derived from NIR and mm data, as well as the myriad non-axisymmetric structures observed in the disk suggest, albeit indirectly, that a massive object or objects may be responsible for carving the transitional disk gap in HD 100546. Indeed, a thermal infrared (\textit{L}'-band, 3.8$\mu$m) planet candidate, HD 100546 \textit{b}, has been detected with adaptive optics observations at $60$ au from the central star several times \citep{Quanz:2013,Quanz:2015,Currie:2015}, although it lies too close to the central star to be responsible for the mm-derived outer disk gap at 190au and too far to be responsible for the cavity interior to $\sim$15au. Subsequent \textit{Ks}-band (2.15$\mu$m) observations of the disk did not reveal a point source at the location of the \textit{b} candidate \citep{boccaletti13}, but rather faint extended emission \citep{Garufi:2016}. The nature of and physical relationship between the more compact \textit{L}' source and the extended \textit{Ks} source is a subject for debate, and we discuss this in more detail in the companion to this paper (Rameau et al., submitted), which is focused on the HD 100546 \textit{b} planet candidate. 

Another candidate object (HD 100546 ``\textit{c}") was also put forward to explain spectroastrometry of CO and OH emission lines in HD 100546 \citep{Brittain:2014}, at a separation of $\sim$15au, just inside of the NIR inner disk rim. However, the planet explanation for the spectroastrometric signature has been called into question by \citet{Fedele:15}. Using the Gemini Planet Imager (GPI, \citealt{Macintosh:2014js}), the direct detection of a second planet candidate (HD 100546 ``\textit{c}") at \textit{H}-band has also been put forward \citep{Currie:2015}, but has yet to be confirmed. 

In this paper, we report observations of HD 100546 obtained with GPI as part of the GPI Exoplanet Survey (GPIES) and with the Magellan Visible AO camera  (VisAO) as part of the Giant Accreting Protoplanet Survey (GAPlanetS, Follette et al. \textit{in prep.}). These high-resolution multiwavelength images reveal fine structures that can be compared to images obtained with other AO instruments and to model images to assess their robustness to various processing techniques. A companion paper will focus on GPI and MagAO derived limits on emission from the planet \textit{b} (Rameau et al., submitted), while this paper will focus on revealed disk structures and limits on planet ``\textit{c}". 

When imaged with adaptive optics systems, point sources such as stars are surrounded by a halo of light from uncorrected or mis-corrected wavefront errors. Instantaneously and monochromatically, this point spread function ﴾PSF﴿ consists of an interference pattern of ``speckles" of size similar to the diffraction limit of the telescope. In long exposures this speckle pattern partially smoothes out as the wavefront changes, but retains some structure on timesceales of minutes or longer (``quasi‐static speckles") due to static optical errors, as well as asymmetries e.g. due to stronger wavefront errors along the direction of wind propagation.

The surface brightness of the HD100546 disk is lower than that of the stellar halo and hence this halo must be removed through PSF subtraction. The application of these algorithms are discussed in detail in Section \ref{sec:hspecdata}.  It is important to note, however, that PSF subtraction algorithms are typically optimized for point source extraction. Many groups have now demonstrated success at extracting disks with these algorithms \citep[e.g.,][]{Milli:2012, Rodigas:2012, Mazoyer:2014, Perrin:2015}, however the majority of successful extractions have been of debris disks with either edge-on or ring-like morphologies. Young, extended disks, and in particular disks with moderate inclination such as HD 100546, are more problematic because their large angular and radial extent means that disk emission at a given location is present in  many (if not most) reference PSFs. This has led some  to question the reality of structures visible after aggressive post-processing \citep[e.g.][]{boccaletti13}.   

GPI and VisAO observations and image processing are described in Section \ref{sec:obs}. Measurements derived from these processed images are presented in Section \ref{sec:results}. Interpretation of these results, as well as a qualitative comparison of our results to planet-driven spiral disk model images processed in a similar manner are discussed in Section \ref{sec:discuss}. We provide conclusions in Section \ref{sec:conclude}. Constraints on the \textit{b} planet candidate are presented in a companion to this paper (Rameau et al., submitted).

\section{Observations and Data Reduction} 
\label{sec:obs}

\subsection{Gemini Planet Imager Data}

Initial Gemini Planet Imager observations of HD 100546 were taken in \textit{H}-band spectroscopic mode (hereafter \textit{H}-spec) using Angular Differential Imaging \citep[ADI,][]{Marois:2006df} as part of the Gemini Planet Imager Exoplanet Survey (GPIES). Followup observations were conducted in both spectroscopic and polarimetric mode based on extended structures suggested by this preliminary dataset. A full summary GPIES observations is given in table \ref{tab:data}. All initial reductions were done using the GPI Data Reduction Pipeline (DRP)  version 1.3.0 \citep{Perrin:2014, Perrin:2016}. We refer the reader to these papers for full details of the GPI DRP. In brief, the GPI DRP subtracts dark background, interpolates over bad pixels, corrects for DC offsets in the 32 readout channels, and converts from raw 2D IFS frames to 3D datacubes. In the case of spectral data, Argon arc lamp exposures taken both at the beginning of the night and immediately prior the science exposure sequence are used for wavelength calibration, and the locations and fluxes of the four satellite spots created by the apodizer are used to compute and apply astrometric and photometric calibrations. In the case of polarimetric data, the pipeline assembles a full Stokes datacube from the sequence of exposures.

\begin{table*} [t]
\centering
\begin{tabular}{ccccccccc}
%\caption{Summary of Gemini and Magellan Datasets}
%\tablecolumns{9}
%\tablehead{
\hline\hline
Instrument & Date & Observing Mode & n$_{\rm images}$ & t$_{\rm int}/$frame & n$_{\rm coadds}$ & t$_{\rm int}$ total & $\theta_{\rm rot}$ & avg. seeing \tablenotemark{a} \\
 & & & & (sec) & & (min) & (deg) & ($\arcsec$) \\
\hline
MagAO & 2014-04-11 & H$\alpha$ SDI & 3423 & 2.273 & 1 & 129.7 & 73.5 \tablenotemark{b} & 1.05 \\
\textbf{MagAO} & \textbf{2014-04-12} & \textbf{H$\alpha$ SDI} & \textbf{4939} & \textbf{2.273} & \textbf{1} & \textbf{187.1} & \textbf{71.6} & \textbf{0.58} \\
GPI & 2014-12-17\tablenotemark{a} & \textit{H}-spec & 33 & 60 & 1 & 33 &  12.9 & \nodata \\
GPI & 2015-01-30\tablenotemark{a} & \textit{Y}-pol & 14 & 60 & 1  & 14 & \nodata & 0.63 \\
MagAO & 2015-05-15 & H$\alpha$ SDI & 2077 & 2.273 & 1 & 78.7 & 42.0 & 0.46 \\
\textbf{GPI} & \textbf{2016-02-27} & \textbf{\textit{H}-spec} & \textbf{120} & \textbf{60} & \textbf{1} & \textbf{120} & \textbf{51.6} & \textbf{0.66} \\
\textbf{GPI} & \textbf{2016-01-28} & \textbf{\textit{Y}-pol} & \textbf{62} & \textbf{15} & \textbf{4} & \textbf{62} & \textbf{\nodata} & \textbf{0.69}  \\
\hline
\end{tabular}
\\
\begin{flushleft}
a. The instrument and method for measuring seeing varies by telescope and observing run. Magellan seeing values are derived from measurements taken at the Baade telescope. Gemini South has both MASS and DIMM seeing monitors, however only the DIMM was functioning on 2014-04-11, 2014-04-12 and 2015-01-30 and neither was functioning on 2014-12-17. On 2015-02-27 and 2016-02-28, both were online and seeing recorded by the two instruments has been averaged. \\
b. Although this dataset has a total of $73.5^{\circ}$ rotation, the space is not evenly sampled and there is a $10^{\circ}$ gap in rotational space while the system was pointed at a NIR reference PSF star. \\
\textbf{Bolded rows} represent the three highest-quality datasets, which are used for the bulk of the analyses in this paper.
\end{flushleft}
\label{tab:data}
\end{table*}

\subsubsection{Polarimetric Data}

\textit{Y}-band polarimetric images (hereafter \textit{Y}-pol) were first attempted on 2015-01-30, however the sequence was aborted due to poor conditions. The dataset we analyze in this work was collected on 2016-01-28. Data were taken using the shortest GPI filter (\textit{Y}-band, 0.95--1.14$\mu$m) and accompanying \textit{Y}-band coronagraph because this mode affords the highest angular resolution and has the smallest coronagraphic inner working angle (0$\farcs$078), allowing us to probe the very inner regions of the disk near the HD 100546 ``\textit{c}" source and inside the inner cavity rim. Standard polarimetric differential imaging waveplate cycles of $0^{\circ}$, $22.5^{\circ}$, $45^{\circ}$ and $67.5^{\circ}$ were employed in order to allow for double difference polarized imaging \citep{Kuhn:01,Perrin:2004, Hinkley:2009, Hashimoto:2011}, and the 60 second images were taken as four 15 second coadds in order to avoid saturating the inner region of the images. 

Generation of the Stokes cubes was done using the standard polarimetry recipes available in the GPI pipeline and described in detail in \citet{Perrin:2014, Millar-Blanchaer:2015}, however three modifications were made to the standard polarization recipe. First, we assumed a perfect half waveplate rather than the default lab-measured waveplate retardance. We have found in some cases that this improves the final image quality, as measured by the amount of residual signal in the $U_r$ image (see next paragraph). The mean stellar polarization was estimated for each datacube individually from the normalized difference of the two orthogonal polarization slices in the region $0<r<5$ pixels ($r<0\farcs07$), which is beneath the focal plane mask. Light that lies in this region should be composed primarily of starlight diffracted around the FPM, and any polarized signal is most likely induced by the instrument optics if we assume that the starlight is intrinsically unpolarized. The mean normalized difference in this region is scaled to the total polarized flux in each pixel before removal.  For more details about the specifics of this estimation, see \citet{Millar-Blanchaer:2016}. Finally, we smoothed the processed images with a 2-pixel FWHM Gaussian kernel before combining into the Stokes cube (I, Q, U, V) in order to mitigate microphonics noise \citep{Ingraham:2014}. 

The Stokes cube generated by the GPI DRP was transformed to a radial Stokes cube ($I$, $Q_{\phi}, U_{\phi}$, $V$; see \citealp{Schmid:2006}) via the same method as in \citet{Millar-Blanchaer:2015}. Under this convention, all polarized signal oriented parallel or perpendicular to the vector connecting the pixel to the central star is encompassed in the $Q_{\phi}$ image, and all signal oriented at $\pm$45$^{\circ}$ is encompassed in the $U_{\phi}$ signal. The $Q_{\phi}$ image thus contains the centrosymmetric polarized disk signal in the case of single-scattering, and the $U_{\phi}$ image is an approximation of the noise, under the assumption the contribution of multiple-scattered photons is small. 

The final $Q_{\phi}$ and $U_{\phi}$ images are shown in Figure \ref{fig:Ypol} and discussed in detail in Section \ref{sec:Ypol}. 

\subsubsection{Spectroscopic Data}
\label{sec:hspecdata}
The \textit{H}-spec coronagraphic dataset taken on 2014-12-27, while nominally a full GPI sequence, had low overall field rotation ($12.9^{\circ}$), and its utility was compromised as a result. For this moderate inclination and highly extended disk, a large amount of field rotation is necessary to minimize disk self-subtraction and extract robust disk structure, therefore we followed up this initial observation on 2016-02-27 with a two hour on sky sequence, reaching $51.6^{\circ}$ field rotation under good weather conditions (see Table \ref{tab:data}). Results based on these later observations are presented in this paper. %We note this dataset achieves twice the rotation and total integration time of the GPI dataset presented in \citet{Currie:2015} (24$^{\circ}$ and 55min, respectively), and therefore offers a significant improvement in our ability to mitigate disk self-subtraction and image faint disk structures.

The GPI DRP processes the raw data through dark subtraction, wavelength calibration based on observations of an argon arc lamp  \citep{Wolff:2014}, bad pixel identification and interpolation, microspectra extraction to create $(x,y,\lambda)$ datacubes \citep{Maire:2014gs}, interpolation  to a common wavelength axis, and distortion correction \citep{Konopacky:2014}. Astrometric calibration (platescale of $14.166\pm0.007$ mas/pixel, position angle offset of $-0.10\pm0.13^{\circ}$) was obtained with observations of the $\theta_1$ Ori field and other calibration binaries following the procedure described in \citet{Konopacky:2014}.

Further post processing was also done using the GPI DRP. The 3-D datacubes were first aligned using the photocenter of the four satellite spot positions \citep{Wang:2014}. To remove slowly evolving large scale structures, the datacubes were high-pass filtered using a smooth Fourier filter with cutoff frequencies between $4$ and $16$ equivalent-pixels in the image framework, allowing us to investigate disk features on different spatial scales. Since this step strongly affects the apparent geometry of the disk, the two extremes of these cutoff frequencies, as well as images without any highpass filter applied are discussed and shown in Section 3.2. 

The stellar point spread function (PSF) was estimated and subtracted from each image in the sequence using several ADI algorithms: classical Angular Differential Imaging (cADI, \citet{Marois:2006df}), Locally Optimized Combinations of Images (LOCI, \citet{Lafreniere:2007}) and Karhunen-Lo\`eve Image Processing (KLIP, a form of Principal Component Analysis, \citet{Amara:2012, Soummer:2012}) via a custom IDL pipeline.  Using different ADI algorithms was mandatory in the analysis of this inclined, asymmetric, bright, and extended transitional disk to better assess the robustness of resolved structures against residual speckles, which manifest themselves differently in the post-processed images computed by each algorithm. For all three algorithms, residual images were rotated to align north with the vertical, combined with a 10$\%$ trimmed mean (discarding the highest and lowest 5$\%$ of pixel values in the temporal sequence), and collapsed over the wavelength axis to create a final broad-band image. 

cADI processing has no tunable parameters. The stellar PSF subtracted from each image is simply the median of the entire image cube, and the PSF-subtracted images are then rotated to a common on-sky orientation before combining. This method therefore is not capable of removing evolving PSF features, but it provides  a good estimate of the most static PSF structures. Though mitigated by the large amount of field rotation, the disk extends azimuthally over more than the 51.6$^{\circ}$ of rotation in the dataset, so some disk emission survives into the median PSF, resulting in negative ``self-subtraction" regions at the edges of the disk. 

LOCI analysis was done with annuli of $dr=5$ pixels, optimization region of $N_A=500$ full width at half maximun (FWHM, $3.6$ pixels at $H$ band), geometry factor $g=1$, and minimum separation criterion $N_\delta=1$ FWHM. 

KLIP analysis was done on a single image region from 5 to 100 pixels in radius (0$\farcs$07 to 1$\farcs$42) and keeping only the first one to five Karhunen-Lo\`eve (KL) modes. Although KLIP is typically used for point-source searches with more zones and a greater number of KL modes, this single zone, small number of KL-mode approach is standard for minimizing self-subtraction of extended disk features. 

The PSF was also subtracted using the Reference Differential Imaging technique (RDI) implemented in the TLOCI quick look processing pipeline (an evolution of the SOSIE pipeline, \citet{Marois:2010b}). A library of reference images was created from 426 \textit{H}-band data cubes (all GPIES campaign observations taken in pupil-stabilized mode at \textit{H}-band at the time of processing). Data from each reference sequence were first reduced with the GPI DRP in the standard manner described previously. Additionally, each image in an object sequence was high pass filtered using an 11 pixel (0$\farcs$16, 4 $\lambda/D$) square unsharp mask, magnified to align speckles across wavelength channels, flux normalized so that the satellite spot intensities were the same in each channel, and wavelength collapsed (only slices 5--31 were used to avoid the noisy wavelength slices at the end of every GPI spectral cube). These high-pass filtered, aligned, normalized and wavelength-collapsed images were then median-combined for each object sequence and scaled to the flux of the target star using the satellite spots, allowing us to gather a homogeneous library of achromatic speckle-limited images with greatly reduced disk, planet or background star signals. The HD 100546 DRP images were processed through TLOCI RDI pipeline using only the 20 most correlated reference images in this PSF library to subtract the speckle noise. Reference images were selected by performing a cross-correlation analysis in a $[15-80]$ pixel (5.4-28.6 $\lambda/D$ , 0$\farcs$212-1$\farcs$133) annular region to avoid the focal plane mask edge. 

RDI, cADI, LOCI and KLIP images are shown in Figure \ref{fig:Hspec} and are discussed in detail in Section \ref{sec:Hspec}. 

\subsection{Magellan VisAO Data}

High contrast, visible light, adaptive optics observations were conducted at the Magellan Clay telescope with the Magellan Adaptive Optics System (MagAO, \citealp{Morzinski:2016, Morzinski:2014}) and its visible light camera (VisAO, \citealp{Males:2014}). The observations were conducted in H$\alpha$ Simultaneous Differential Imaging (SDI) mode in which a Wollaston prism is used to split the beam into two channels, and each is passed through a separate narrowband filter, one centered on the H$\alpha$ emission line (656 nm, $\Delta\lambda$=6 nm) and one centered on the nearby continuum (642 nm, $\Delta\lambda$=6 nm ). The continuum image serves as a sensitive and simultaneous probe of the stellar PSF. 

GAPlanetS data are reduced with a custom IDL pipeline. Raw data frames are bias subtracted and divided by a flat field generated from \textit{R}-band twilight sky observations. Dust spots in the instrument optics create significant throughput effects and can create point-like artifacts, but are clearly revealed in the flat field. They are not effectively removed by simply dividing by the flat field (why is unclear), and we therefore mask all pixels within 2 pixels of a region with $<$98\%  throughput. This mask is applied to all data frames before further analysis. 

The bias-subtracted and flat fielded raw images are then separated into line (e.g., H$\alpha$) and continuum channels. Individual channel images are registered against a high quality individual data frame. The center of rotation is found through a custom centering algorithm  that locates the center of circular symmetry in the median collapsed registered datacube by finding the point that minimizes the standard deviation of intensity in annuli centered at that point. We find that this algorithm performs better than radon transform or center of rotational symmetry algorithms for VisAO data, as measured using a binary of well-known separation and PA. 

Although a minimum integration time of 2.273 seconds was used in all cases, the HD 100546 observations were saturated at radii interior to 7 pixels in all datasets. This was noted during observations, however we opted to saturate the very inner region rather than decrease detector gains and therefore sensitivity. We apply a software mask of $r=8$ pixels to all data before centering, and exclude pixels interior to this radius in our KLIP reductions. 

\begin{figure*}
\centering
\includegraphics[width=7in]{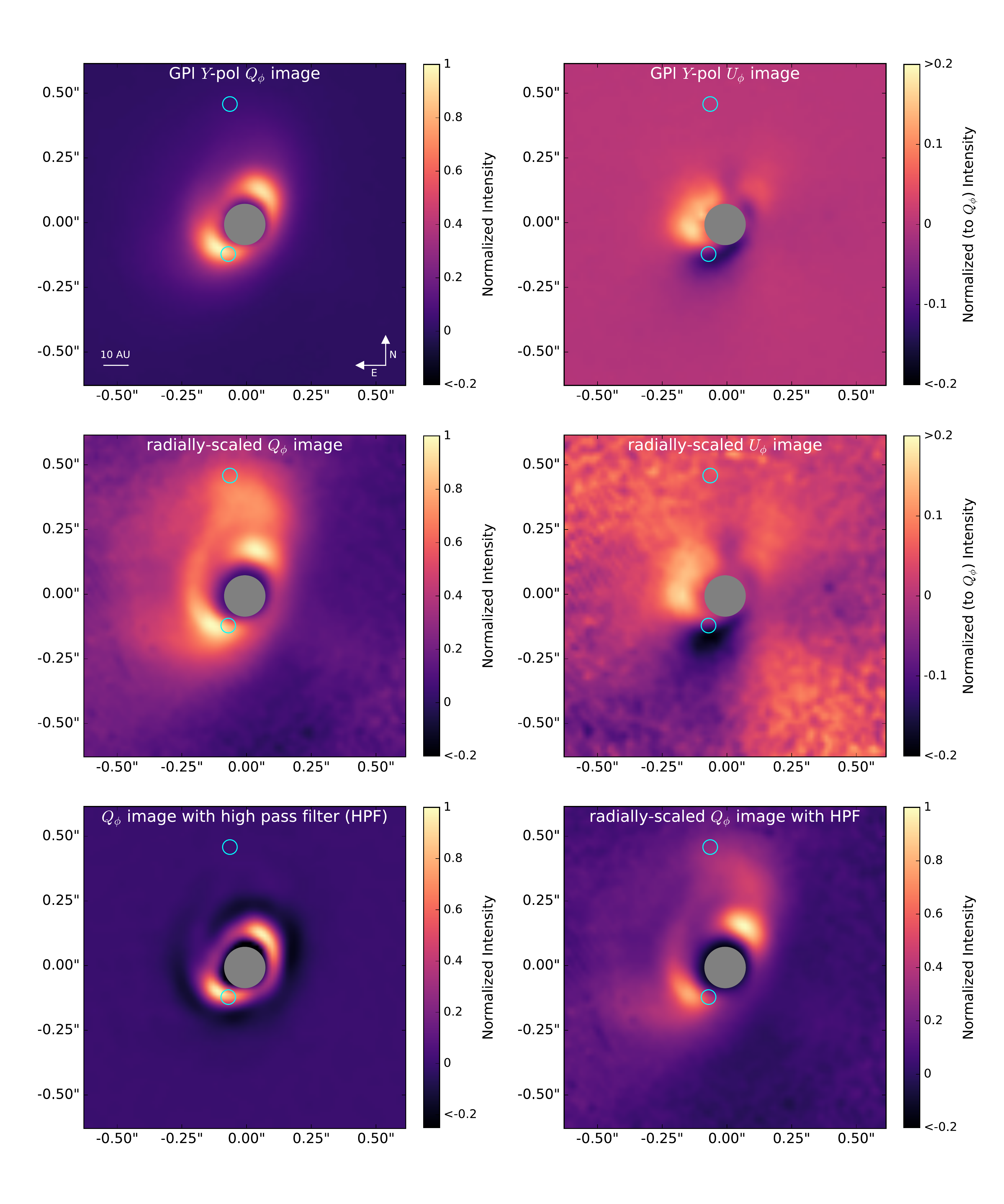}
\caption{GPI Y band radial polarized intensity ($Q_R$) images. top left: The GPI $Q_{\phi}$ image, top right: The GPI $U_{\phi}$ image, normalized relative to the peak value of the $Q_{\phi}$ image and shown with a tighter stretch so that the structures are visible. middle left: The $Q_{\phi}$ image scaled by $r^{2}$ for a disk inclined at 42$^{\circ}$ along a PA of 145$^{\circ}$, middle right: The same $r^{2}$ scaling applied to the $U_{\phi}$ image. lower left: The $Q_{\phi}$ image with a 4 pixel Fourier high-pass filter applied, and lower right: The $r^{2}$ scaled $Q_{\phi}$ image with a 4 pixel Fourier high-pass filter applied. The northeastern spiral is readily apparent extending from the eastern disk rim toward the north in all but the unaltered $Q_{\phi}$ image. Cyan circles indicate the locations of the candidate ``b" and ``c" protoplanets, and the grey circles indicate the GPI $Y$ band coronagraph occulter. All images have been normalized by dividing by the peak pixel value.} 
\label{fig:Ypol}
\end{figure*}

\begin{figure*}
\centering
\includegraphics[width=7in]{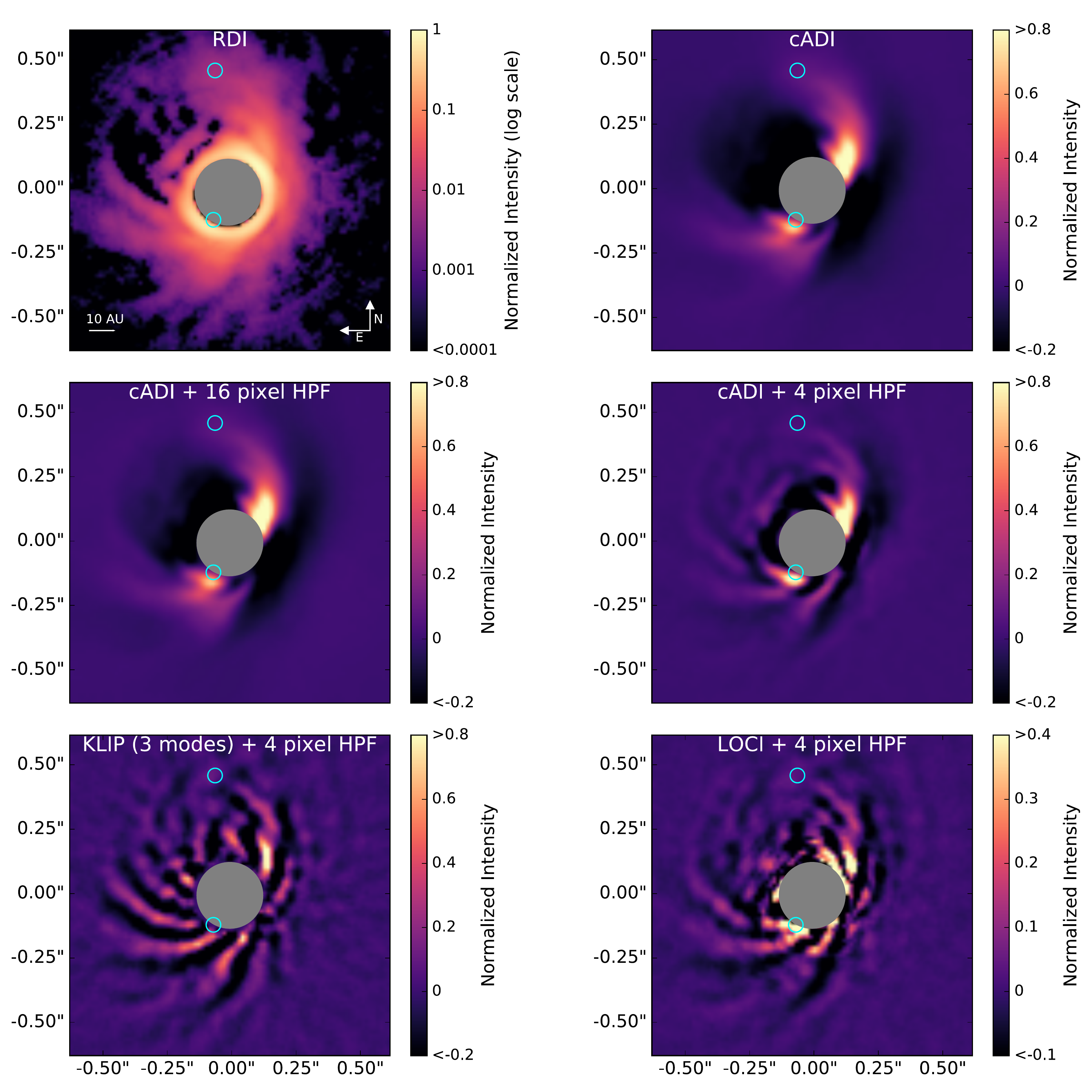}
\caption{GPI \textit{H}-band total intensity images of HD 100546 using different algorithms. The reduction algorithms increase in aggressiveness from top to bottom, and are described in detail in the text. The locations of the candidate protoplanets ``b" and ``c" are marked with cyan circles. All images have been normalized by dividing by the peak pixel value. The RDI image has been log-scaled to reveal faint outer disk structures, but this is impractical for the other images, which have large self-subtraction regions.}
\label{fig:Hspec}
\end{figure*}

\begin{figure*}
\centering
\includegraphics[width=7in]{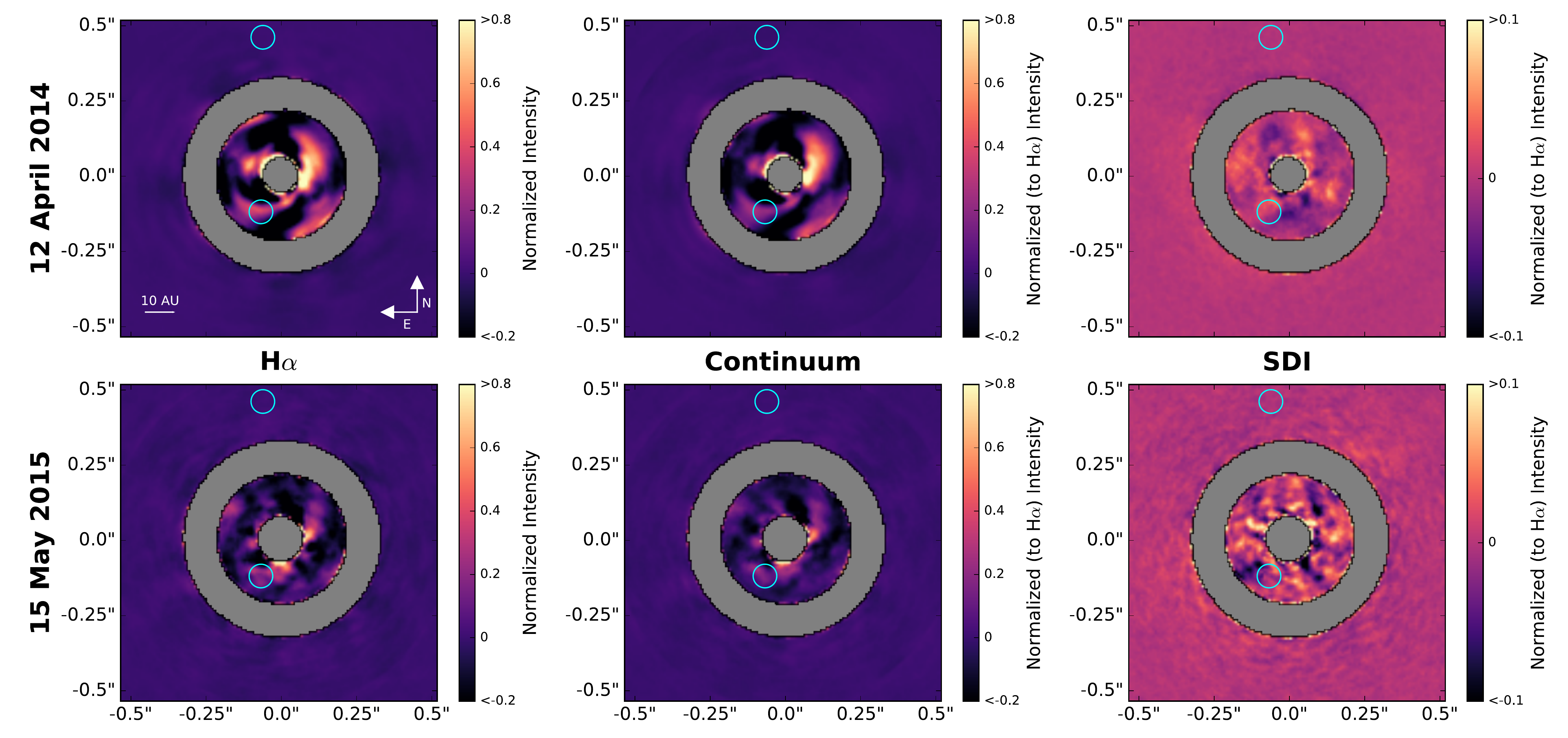}
\caption{MagAO H$\alpha$ SDI images of HD 100546 from 2014-04-12 (top) and 2015-05-15 (bottom). The H$\alpha$ (left panels) images are dominated by scattered light structures at or near the disk rim. These features are closely mimicked in the continuum images (middle panel), albeit at slightly lower intensity due to stellar H$\alpha$ excess and their scattered light nature. The rightmost panels represent the SDI images for each dataset, generated by scaling and subtracting the continuum images from the H$\alpha$ images and combining. No H$\alpha$ excess sources are visible in either SDI image, including at the locations of the HD 100546 ``b" and ``c" planet candidates. The region surrounding the AO control radius, where spurious speckle structures dominate KLIP reductions, has been masked in all images. All images have been normalized by dividing by the peak pixel value in the H$\alpha$ image for that epoch.}
\label{fig:magao}
\end{figure*}

It is important to note that although H$\alpha$ emission is typically thought of as an accretion tracer, disk-scattered light also makes a significant contribution at this wavelength, particularly in cases where the star itself is actively accreting. In fact, disk-scattered light is ubiquitous even in PSF-subtracted images because of the extended and moderately-inclined nature of the HD 100546 disk. The MagAO system  was designed to utilize the simultaneous nature of our H$\alpha$ and continuum observations to remove both direct starlight and disk-scattered light contributions that are equivalent in the two filters. We compensate for the difference in stellar (and therefore scattered light) brightness between the two filters by subtracting a scaled version of the simultaneous continuum image from each H$\alpha$ image before further processing. The scaling factor is determined iteratively as the value that results in minimized noise residuals in the region $8<r<27$ pixels (representing the region between where the saturated images reach linearity and the inner boundary of the AO control radius). Scaling and subtracting the continuum image in this way effectively removes the contribution of scattered light disk structures and diffracted starlight from the images. Because accreting protoplanets are expected to exhibit H$\alpha$ excess and to not have a detectable level of continuum emission, this strategy should eliminate starlight and disk scattered light preferentially, leaving behind pure H$\alpha$ emission.It is the KLIP-processed versions of these SDI images that we use to place constraints on H$\alpha$ emission from accreting protoplanets in these datasets. 

KLIP images are generated using the MagAO interface of \texttt{pyKLIP}, a Python implementation of the KLIP algorithm \citep{Wang:2015}. Of particular importance to the discussion in this paper is the fact that the final images are very sensitive to our choice of KLIP parameters, notably zone size and masking parameters. Although not exhaustive, we explore a wide region of this parameter space in order to assess the robustness of the parameters we extract, as reported in Section \ref{sec:results}. The AO control radius for the MagAO system lies at $r=35$ pixels in our images, and we find that the region $27<r<42$ pixels is particularly noisy as a result, with many short-lived speckles that are not well subtracted with KLIP. We therefore mask this region in each image before KLIP processing to avoid the appearance of spurious structures in the final reductions.

The moderate inclination of the disk means that disk structures cover wide swaths in azimuth, and aggressive KLIP reductions can be potentially problematic. We find that KLIP reductions with small to moderate exclusion criteria (e.g., allowing images where a planet would have moved by fewer than 8 pixels within a given annulus) result in large heavily self-subtracted regions and turn extended disk features into spurious point-sources. 

We elect the least-aggressive exclusion criterion possible for each dataset, excluding all images from the reference library where a hypothetical planet located in the center of an annulus would have moved by fewer than a given number of pixels, where that number is as large as possible. For the 2014 dataset, this is 12 pixels, corresponding to 33 degrees of rotation in the innermost annulus before an image is included in the reference library. For the 2015 dataset, this is 8 pixels, corresponding to 21 degrees of rotation.  Since the maximum exclusion criterion is nearly twice as aggressive in the case of the 2015 dataset, it is unsurprising that the disk rim is not as cleanly revealed as in the 2014 reductions. 

KLIP-processed H$\alpha$, Continuum and SDI images for both epochs are shown in Figure \ref{fig:magao} and discussed in detail in Section \ref{sec:magao}.

\section{Results}
\label{sec:results}

\subsection{GPI \textit{Y}-band Polarimetric Imagery}
\label{sec:Ypol}

GPI \textit{Y}-pol images, shown in Figure \ref{fig:Ypol} clearly resolve the scattered light cavity rim. There are distinct bright lobes along the disk major axis, however these are symmetric about the star, and we see no evidence in these data of anything unusual at the location of the purported HD 100546 ``\textit{c}" point source.

The corresponding GPI $U_{\phi}$ image shows a non-zero signal, peaking at $\sim$20\% of the value of the $Q_{\phi}$ image with most of the signal localized east of the star and just outside the coronagraph. This is potentially an effect of instrumental polarization, but non-zero signal in $U_{\phi}$ HD100546 images has been seen before \citep[albeit with a different signal morphology,][]{avenhaus14,Garufi:2016}, and may be a result of physical rather than instrumental effects. For example, multiple-scattering is expected to create non-zero $U_{\phi}$ signals \citep{Canovas:2015}.

In order to compensate for the purely geometric $r^{-2}$ dropoff in stellar scattered light, we scaled the images by $r^{2}$ for a disk inclined at 42$^{\circ}$ along a PA of 145$^{\circ}$, a common practice in the field for revealing fainter extended structures in the outer disk. We note that we apply this scaling only to highlight faint disk features and that any asymmetries in brightness or location of disk features along the minor axis are impacted by the inclined, vertically-extended and optically thick nature of the disk, which will tend to artificially enhance the illuminated half of the disk. The $r^{2}$-scaled images do, however, effectively reveal a faint extended feature connected to the southeastern disk rim and extending to the north, which we will refer to hereafter as the ``northeastern spiral". This feature is also effectively revealed with a simple 4 pixel Fourier high-pass filtering of the original image. This and other morphological features revealed in GPI and MagAO imagery are discussed in detail in Section \ref{sec:features}. 

\begin{figure}
\centering
\includegraphics[width=3.5in]{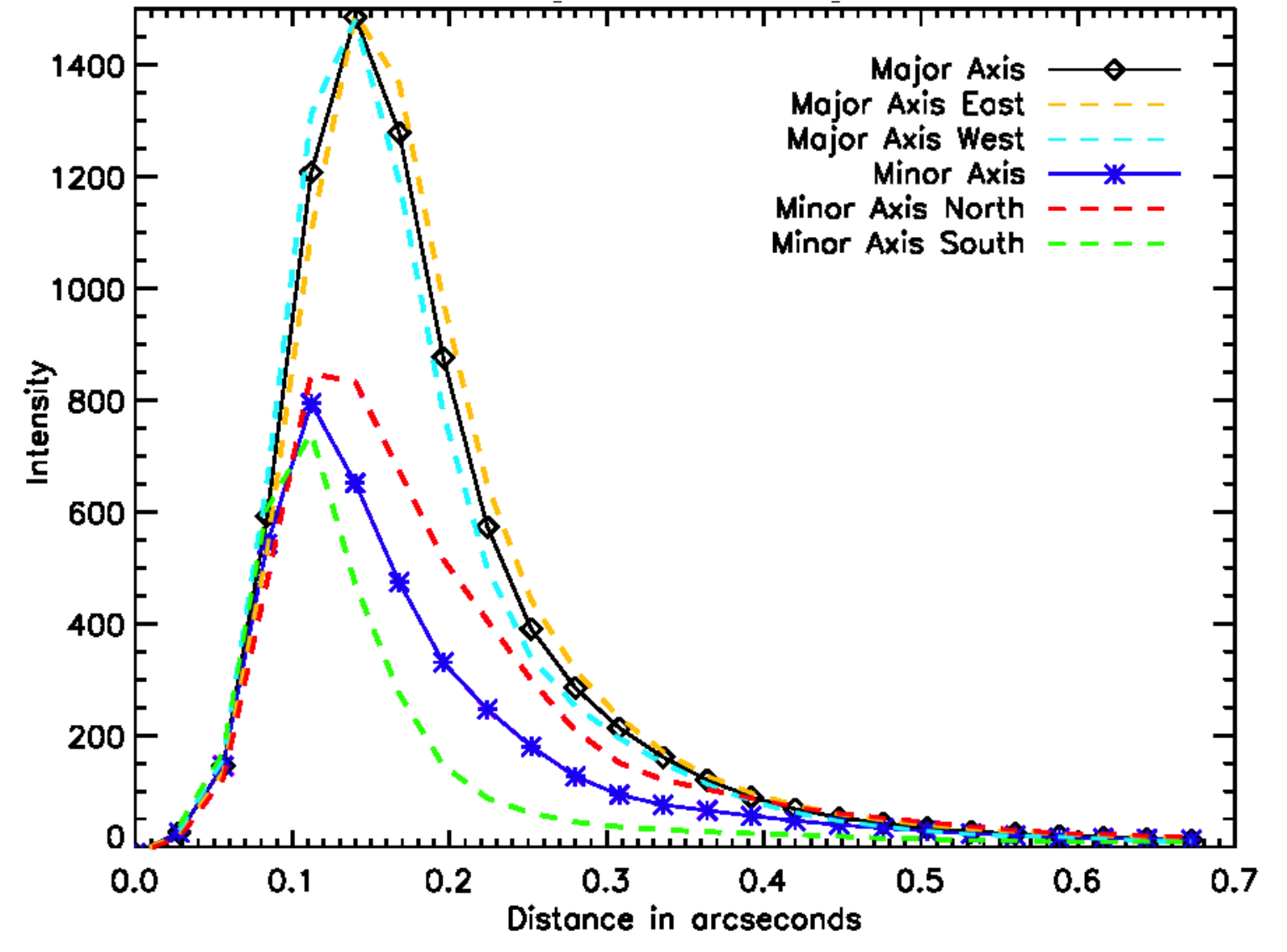}
\caption{Radial profiles for \textit{Y}-band radial polarization image along the major (black diamonds) and minor (blue stars) axes. In each case, the profile is averaged across the two sides of the disk, but the individual profiles are also shown as dashed lines to assess symmetry. The eastern and western major axis profiles are virtually identical, suggesting that there is no significant asymmetry in the peak brightness or the location of the disk rim along the major axis. The brighter and more distant peak of the northern minor axis profile relative to the southern is expected given the north part is the illuminated half of the disk, as explained in the text.  \label{fig:Yrp}}
\end{figure}

Radial profiles taken through the GPI Y-pol images, shown in Figure \ref{fig:Yrp}, reveal that there is no significant deviation between profiles taken to the east and west along the major axis, despite the proposed existence of a planet candidate along the Eastern major axis. The profiles peak at 0$\farcs$14, suggesting a cavity rim at 15au. This is marginally inconsistent with the cavity radius estimated with SPHERE at \textit{R} of 12.5$\pm$1 au, but quite consistent with the range of estimates (15-17au) in the literature for the NIR cavity rim. 

The minor axis profiles are significantly different both in radial extent and in absolute intensity along the Northern and Southern minor axis, however this is an expected effect. The greater radial extent and brightness of the northern minor axis profile is consistent with that being the illuminated half of the disk, and is likely affected by both the geometry of the disk and the scattering phase function.

Given the dearth of successful detections of polarized light from young planets in the literature (only upper limits e.g., \citet{Jensen-Clem:2016}), it is perhaps unsurprising that there is no evidence of a point source at the location of HD 100546 ``\textit{c}" in the Y-pol image, however there is clear polarized disk structure at this location, and its smoothness and symmetry with respect to disk features opposite the star are surprising in the context of a planet at or near this location.  

\subsection{GPI \textit{H}-band Spectroscopic Imagery}
\label{sec:Hspec}

PSF-subtracted \textit{H}-band images processed through a variety of reduction techniques are shown in Figure \ref{fig:Hspec}, and these techniques increase in aggressiveness toward the bottom of the figure. The apparent morphology is somewhat sensitive to the image processing technique. In particular, less aggressive PSF subtraction techniques (RDI, cADI) result in images that are dominated by an arc of emission extending from SE to NW.  A number of additional, fainter structures resembling spiral arms are present to the South and East of the star, including several in the RDI and cADI images. Aggressive processing with LOCI and KLIP highlights these features further and reveals additional fainter structures, however these aggressive techniques suppress the more extended arc of emission apparent in the cADI and RDI reductions. 

\subsection{MagAO H$\alpha$ SDI Imagery}
\label{sec:magao}

MagAO images are shown for both the H$\alpha$ and continuum channels, as well as SDI images (H$\alpha$ - scale$\times$continuum) in Figure \ref{fig:magao}. The structures in processed continuum images closely mimic the structures in the H$\alpha$ images, which point to their common origin as disk-scattered light. Both images reveal an arc of emission consistent with the forward-scattering portion of the disk rim. 

The importance of field rotation to identification of high-fidelity disk features is apparent in the 15 May 2015 images, which had significantly less field rotation (42.0$^{\circ}$) than the 12 April 2014 images (71.6$^{\circ})$. The same forward-scattering inner disk rim is seen in this case, but it appears clumpy, and structures along it might even be mistaken for point sources. 

The SDI images for both datasets, on the other hand, are free of extended scattered light structures. This points to the effectiveness of the process of scaling and subtracting the continuum image before KLIP processing. The images are also, unfortunately, free of any H$\alpha$ excess point source candidates. This is perhaps unsurprising at the location of the \textit{b} candidate, as it is embedded in the disk and very little dusty material is needed to extinct visible light emission. However, it is somewhat surprising at the location of the ``\textit{c}" planet candidate, which should be minimally extincted if it lies interior to the disk rim and inside of the relatively dust-free disk cavity. Quantitative constraints on detectable contrast levels for the ``\textit{c}" planet are discussed in Section \ref{sec:cplanet}.

It is important to note that the MagAO images presented here have markedly lower Strehl ratios than the GPI images, due to the fact that adaptive optics correction is significantly more difficult to accomplish in the visible than in the NIR since a given optical path difference will correspond to a larger fraction of a wavelength in visible light and naturally produce lower Strehl ratios. How much lower is difficult to estimate given the difficulty of measuring Strehl ratios in general and in saturated data in particular, but they are on the order of $\sim$10--20\% with MagAO at H$\alpha$, $\sim$25--35\% for the GPI \textit{Y}-pol dataset, and $\sim$65--75\% for the GPI \textit{H}-spec dataset. At the same time, the MagAO images benefit from the higher resolution afforded by visible light imaging, which compensates in part for the lower Strehl imagery. 

\begin{figure*}
\includegraphics[width=6.5in]{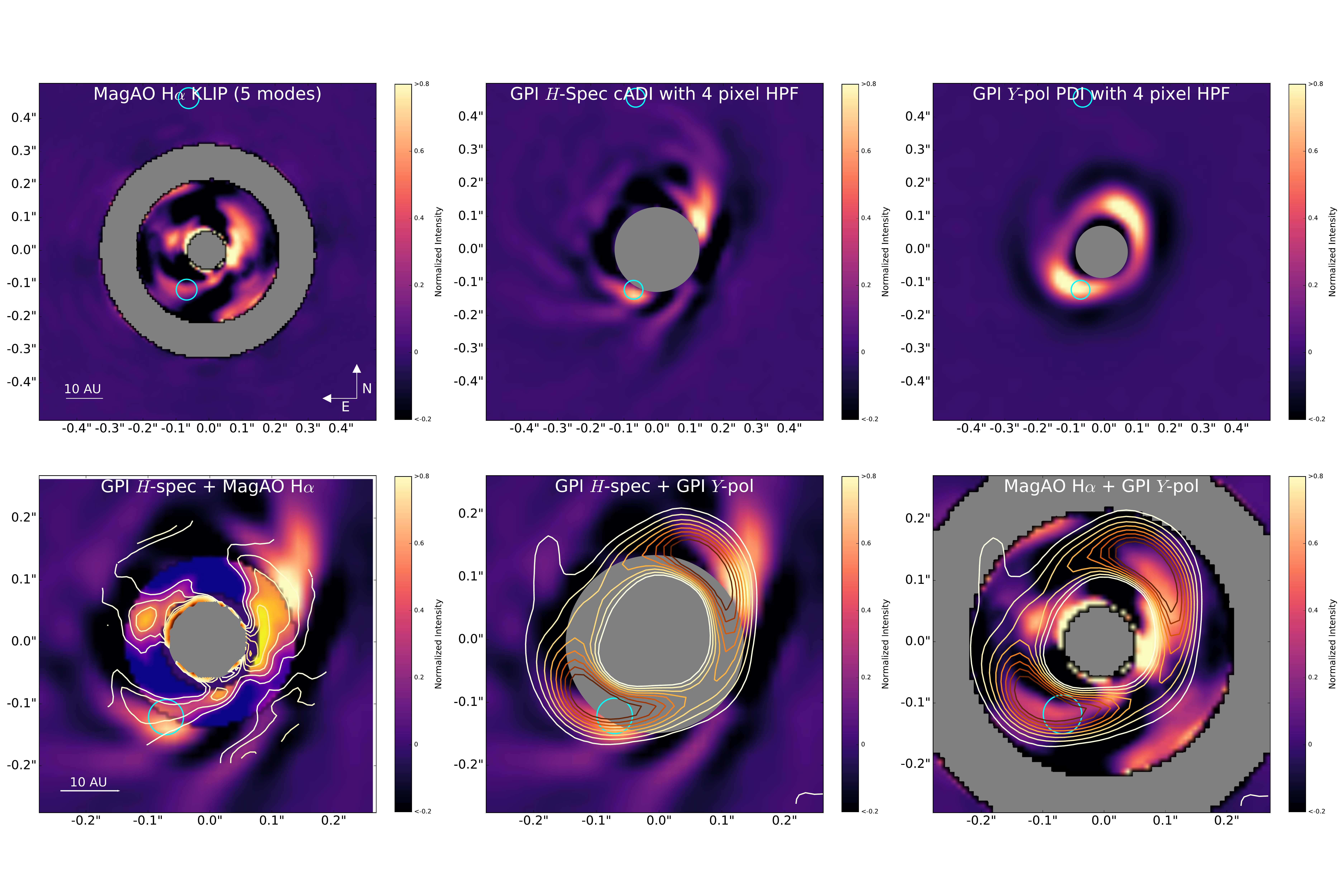}
\caption{ \textbf{Top panels} (left to right): MagAO H$\alpha$, GPI \textit{H}-band total intensity and GPI \textit{Y}-band $Q_R$ polarimetric image of HD 100546 on the same physical scale. The MagAO data were processed using KLIP with 5 KL-modes and parameters as described in the text, and the unreliable saturated and control radius regions are masked in grey. The GPI \textit{H}-spec data have been broadband collapsed, combined via classical Angular Differential Imaging, and processed with a 4 pixel Fourier high-pass filter to reveal the sharper disk structures. The \textit{H}-band coronagrapic mask is shown in grey. The GPI \textit{Y}-band polarized differential image was processed with a 4 pixel high-pass Fourier filter to reveal the northeastern spiral arm. \textbf{Bottom panels}: Zoomed overlays of the images in the upper panels to allow for feature comparisons. \textit{Left}: MagAO contours overlain on GPI \textit{H}-spec data reveal that the Southern spiral arm is contiguous between the two datasets. The region of the MagAO image that lies inside the GPI coronagraphic mask is shown with a different colorscale.  \textit{Middle}: GPI \textit{Y}-pol contours overlain on the GPI \textit{H}-spec image show that the innermost arc of emission in the \textit{H}-spec data is coincident with the disk rim and that the arc of emission stretching to the Northeast in the \textit{H}-spec data is coincident with the northeastern spiral of the \textit{Y}-pol data. \textit{Right}: GPI \textit{Y}-pol contours overlain on the MagAO H$\alpha$ image. \label{fig:overlay}}
\end{figure*}

\section{Discussion}
\label{sec:discuss}

\subsection{Multiwavelength Features}
\label{sec:features}

\begin{figure}[ht]
\centering
\includegraphics[width=0.5\textwidth]{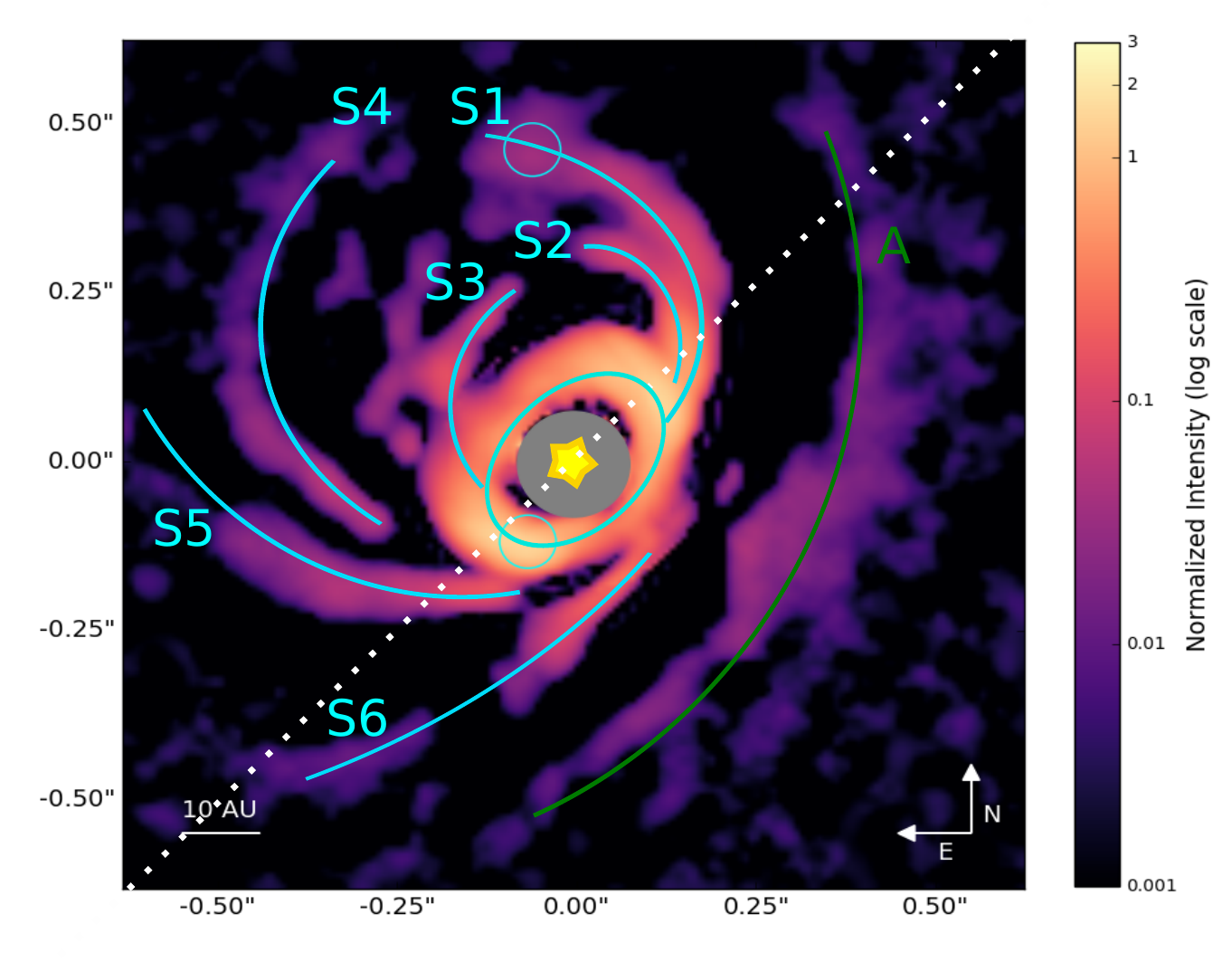}
\caption{A sum of GPI \textit{Y}-pol data (with 4 pixel high-pass Fourier filter), GPI \textit{H}-spec data (cADI with 4 pixel high-pass Fourier filter) and MagAO H$\alpha$ data (from 2014-04-12) HD 100546 datasets. Each image was normalized by dividing by the peak pixel value before summation. Identified features are labeled with aqua (S1--6) and green (A) lines while the dotted white line indicate the disk major axis. \label{fig:schematic}}
\end{figure}

In the previous section, we discussed features revealed in each dataset individually. Here, we discuss how these multiwavelength data complement one another. With the exception of polarized data, where post-processing is minimal, it is unclear from a single dataset alone whether all apparent disk features are true disk structures or artifacts of overly-aggressive PSF subtraction processes. Such techniques have two problems when applied to disks in general, and moderately inclined disks for which significant disk emission survives into reference PSFs in particular. First, surface brightness measurements are severely complicated by disk self-subtraction \citep{Milli:2012}, and we therefore do not attempt them in this work. Secondly, the morphology of complex disk structures can be compromised and spurious point-like artifacts introduced by self-subtraction. By overlaying the three datasets we've obtained and comparing them with features identified previously in the literature, we attempt to address this second point and identify the most robust disk features. 

All three datasets can be seen on the same angular scale in the top panel of Figure \ref{fig:overlay}, and the bottom panel shows pairs of images overlain on one another. The smaller coronagraphic mask in the GPI \textit{Y}-pol dataset and the non-coronagraphic MagAO data allow us to fill in features in the very inner disk region, and the higher sensitivity of the \textit{H}-spec data allows us to probe features in the outer disk. We have selected the \textit{H}-spec cADI processed dataset with a 4 pixel highpass filter for this analysis as it is less-aggressive than the KLIP and LOCI-processed images, but reveals more of the faint disk features than the other cADI images (the highpass filter serves to sharpen the disk features and therefore mitigates the azimuthal extent of the self-subtraction). The overlays reveal several very robust features present in multiple datasets, including the inner disk rim and the northeastern spiral arm. 

We label the most prominent revealed features from all three datasets in Figure \ref{fig:schematic} and discuss them below. We aim here simply to identify and name the most robust features and to compare them to features previously identified in the literature. A detailed discussion of the physical nature of these features, and the spiral arms in particular, is beyond the scope of this work, though we do engage in a brief qualitative comparison with spiral disk models viewed at moderate inclination in Section \ref{modeling}.

\paragraph{Global Near/Far Side Asymmetry} The near side of the disk (inclined toward observer, here the SW side) appears mostly featureless in all three images, whereas most of the structures are present on the far side (NE). This is a natural effect of observing an inclined flared disk, wherein the near side disk geometry causes surface features to be compressed in projection or even shadowed from view by the disk midplane. The \textit{H}-spec data also reveal a bright lane to the southwest, indicated with an ``A" in Figure \ref{fig:schematic}. This feature may be the front edge of the bottom (opposite the disk midplane) side of the disk, as discussed in Section \ref{modeling}. A similarly offset  bright lane feature was recently detected by \citet{deBoer:2016} in the disk of RXJ1615.3-3255 (Feature A1).

\paragraph{Inner cavity} The inner cavity rim seen in both our GPI \textit{Y}-pol data (Figure \ref{fig:Ypol}) and MagAO data (Figure \ref{fig:magao}) and indicated with a cyan ellipse in Figure \ref{fig:schematic} is extremely robust. Its existence is consistent with the NIR deficit in the SED of HD 100546 and with previous resolved images with VLT/NaCo \citep{avenhaus14} and VLT/SPHERE-ZIMPOL \citep{Garufi:2016}, though its location in the \textit{Y}-pol radial profiles is marginally inconsistent with the latter. The potential for disk-self subtraction to affect the apparent location of the disk rim, as well as the close proximity to the \textit{H}-band coronagraph preclude robust measurement of the disk rim location in total intensity at \textit{H}-band or H$\alpha$. Therefore, we defer discussion of whether the marginal inconsistency of our \textit{Y}-pol disk rim radius with the shorter-wavelength SPHERE data is a wavelength-dependent effect for future work. 

\paragraph{Disk ``Wings"} All three of our datasets also reveal an extended arc of emission that runs through and beyond the southern rim of the disk cavity. With aggressive processing, this rim feature can appear sharp, but less aggressive subtractions suggest that it is in fact quite extended. It coincides with the sharp features labeled S5 and S1 in Figure \ref{fig:schematic}, but can best be seen in its extended form in the cADI and RDI images of Figure \ref{fig:Hspec}. It is unclear whether the sharper features that we have labeled S5 and S1 are spirals embedded in that bright wing of emission or are that same feature made sharper by ADI processing. These ``wing" features are the brightest and most distinct features far from the star, and have been identified in several previous studies \citep{Currie:2014,Currie:2015,Garufi:2016}. 

\paragraph{Spiral Arms} The spiral feature labeled S3 in Figure \ref{fig:schematic} is clearly visible in the minimally processed \textit{Y}-pol data, and this also coincides with a brighter region in the Magellan data, though only a portion of it is visible inside of the masked AO control radius region, as revealed by the lower right panel in Figure \ref{fig:overlay}. The \textit{Y}-pol structure is also contiguous with \textit{H}-spec emission that curves toward the feature labeled S2, and it is likely that these two features are part of the same spiral arm. This S3-S2 arm was also seen, though similarly broken, in the deep SPHERE/ZIMPOL polarimetric imagery reported in \citet{Garufi:2016} and by \citet{avenhaus14}. The geometry of this feature is puzzling if it is contiguous, as the apparent curvature back towards the star would suggest that S2 is at least in part a near side feature, yet it does not obscure the cavity. Future deep polarimetric imaging is needed to understand the nature of this feature.

The inner parts of the S1 and S5 features are coincident with the disk ``wings" described above, but the S1 feature curves inward more sharply and is consistent with the ``Northern arm" identified in \citet{Garufi:2016}. It may be contiguous with the feature labeled S4, though, like the apparent S3-S2 spiral, this S4-S1 spiral is broken. The S4 spiral feature is faint and lies in  a region near the bright disk wings that is especially heavily affected by disk self-subtraction, but it too has been seen in previous imagery and is labeled ``spiral 2" in \citet{Currie:2015}.  

The spiral feature S6 is also apparent in both MagAO and GPI \textit{H}-spec data, though there is a break in the revealed feature approximately midway along the line labeled S6. This is the only such feature present on the near (SW) side of the disk major axis in our data. It may be a continuation of a spiral originating on the other side of the disk (S3/2 or S4/1), or it may be a secondary spiral arm mirroring a northern spiral. Similar ``Southern Spirals" were identified in \citet{Garufi:2016}, albeit farther out. The \citet{Garufi:2016} SPHERE \textit{K}-band total intensity images reveal the same feature we have identified as S6, though it is not labeled by the authors as a feature of particular note.

We engaged in a brief exploratory modeling effort, described in Section \ref{modeling}, in an attempt to understand the identified disk structures and the effect that PSF processing can have on them. However, much work remains to be done in this area.

\begin{figure*}[th]
\centering%	
\includegraphics[width=7in]{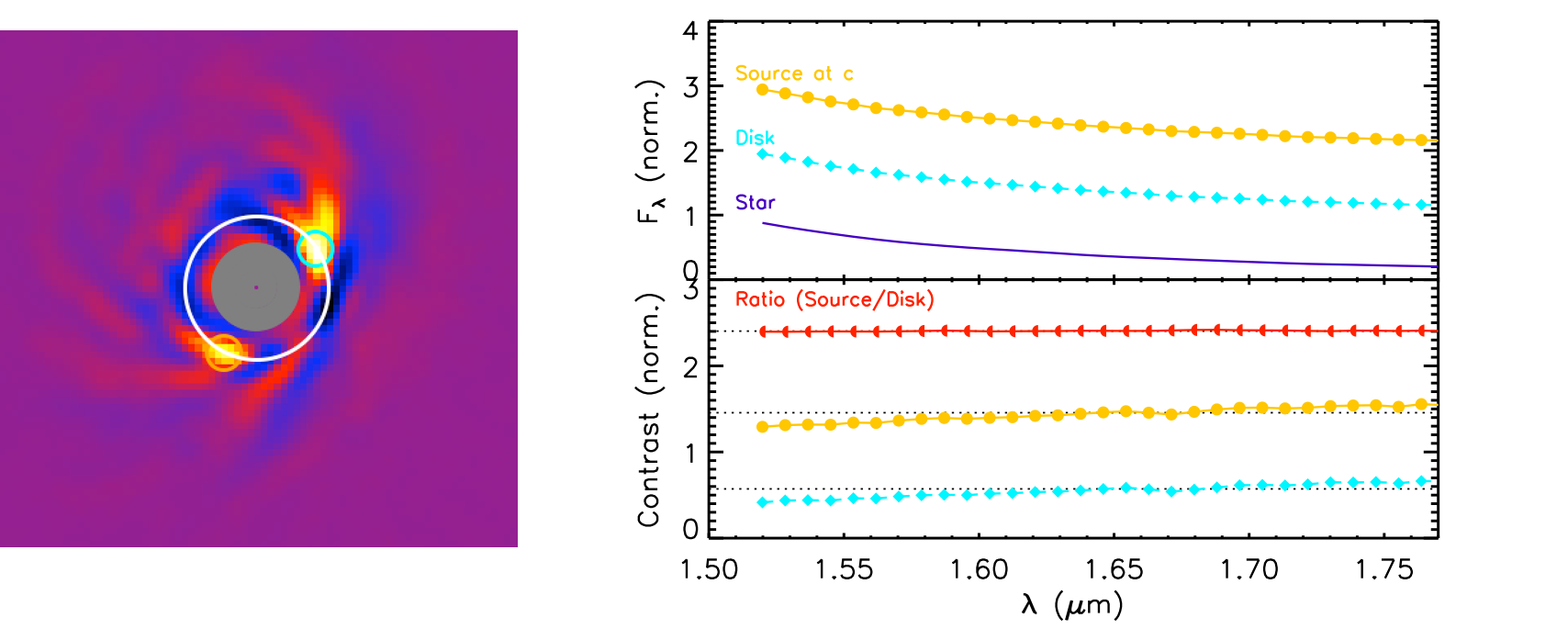}
\caption{(Left) GPI \textit{H}-band residual image after high-pass filter ($4$ pixels) and PCA (KL=1) showing a knot southeast of the star, previously identified as a ``\textit{c}" protoplanet candidate and its symmetric disk counterpart on the opposite side of the disk minor axis. The central region corresponds to a software mask. (Right) Corresponding normalized \textit{H}-band spectra of the two knots (``\textit{c}": yellow circle, disk: aqua circle) and that of the star (purple line) using a BT-NextGen model at $10,500$K \citep{Allard:2009}. Contrast of the extracted ``c" spectrum with respect to the star is plotted in the bottom panel, as is the ratio between the knot at ``\textit{c}" and the symmetric disk knot on the opposite side of the minor axis. Contrasts and spectra are normalized by their mean value and a constant is added to impose an offset for ease of comparison. Since only the relative comparisons were of interest, errors were not computed.}
\label{fig:gpic}
\end{figure*}

\subsection{Limits on the HD 100546 ``\textit{c}" Planet Candidate}
\label{sec:cplanet}

Our MagAO H$\alpha$ and continuum data and GPI \textit{H}-spec data reveal a bright apparent point source at $r=145$ mas and PA=$152^{\circ}$ after aggressive PSF subtraction and/or aggressive high-pass filtering (see Figure \ref{fig:gpic}). This is consistent with the location of the candidate protoplanet put forward by \citet{Brittain:2014} and supported by analyses in \citet{Currie:2015} using a previous GPI \textit{H}-band dataset, so appears at first glance to be a promising planet candidate detection. 

However, it can be seen that this apparent point source is located at the intersection of the disk rim with the northeastern spiral arm and is mirrored by another concentrated knot of emission on the opposite side of the major axis. Although the symmetry of these features and coincidence with spiral arm intersections do not definitively rule out the existence of an underlying point source at the location of the ``c" candidate, they do raise questions regarding its nature. The discovery paper by \citet{Currie:2015} allowed for the possibility that this feature is a disk artifact, and we explore that scenario in this section. 

To assess the hypothesis that the ``c" candidate is a disk artifact, we engaged in two lines of inquiry. 

\paragraph{GPI Spectra} The contrasts of two knots of emission (indicated with circles in Figure \ref{fig:gpic}), one at the location of the ``\textit{c}" candidate and the other at the same location on the opposite side of the star, were extracted from our GPI \textit{H}-band data using aperture photometry with a radius of $0.75\times $FWHM ($3.6$ pixels) using the $4$ pixel high pass filtered PCA (KL=1) reduced wavelength images. Spectra of these knots were obtained after normalization with the spectrum of the star, obtained from the average of a $10,400$ K and $10,600$ K BT-NextGen models \citep{Allard:2009} and binned to the resolution of GPI. Since the two knots lie at the same stellocentric separation, they suffer from equivalent self-subtraction due to ADI and so have the same approximate uncertainties. Since we were only interested in the ratio of the two spectra, we rely on this symmetry to cancel out systematics due to PSF subtraction processing. Results are shown in Figure \ref{fig:gpic} (right panel). Not only does the spectrum of the source at the location of candidate ``\textit{c}" closely match the spectrum of the opposing knot of emission, it also shows no significant deviation from the spectrum of the star, pointing to a scattered light disk origin and showing no indication of an underlying planetary photosphere. 

\paragraph{MagAO SDI imagery} If the ``c" candidate were indeed a protoplanet lying inside the disk gap, we might expect it to be actively accreting as gas passes through the dust cavity en route to the still-accreting central star. The cavity is also depleted in small dust grains, and therefore any H$\alpha$ emission from such an accreting protoplanet should be minimally extincted. Indeed, detecting actively accreting protoplanets through H$\alpha$ emission is the primary motivation behind the GAPlanetS campaign, and this method has been successful twice before \citep{Sallum:2015, Close:2014}. 

Certain aggressive KLIP reductions of the 12 April 2014 MagAO data also reveal a point source candidate at the location of HD 100546 ``c", however a similar point source is also present in the continuum image in all cases, which makes the H$\alpha$ point source immediately suspect, as we do not expect any significant continuum contribution from a substellar object. Scattered light, on the other hand, should appear the same in H$\alpha$ and the continuum and, upon correcting for the H$\alpha$ excess of the primary star (the source of the light to be scattered), should be fully removed by the SDI process. Indeed, as the SDI processed images for both datasets reveal, there is no excess in the H$\alpha$ channel at this location.

In fact, the MagAO images shown in Figure \ref{fig:magao} provide an excellent demonstration of the effects of aggressive PSF processing on extended disk structures. There is significantly less rotation in the 15 May 2015 dataset than in the 12 April 2014 dataset, making the PSF-exclusion criterion necessarily more aggressive (smaller). As a result, structures that appear smooth and extended in the upper panel of the figure appear clumpy and in some cases point-like in the lower panel.  

Taken together, these two lines of evidence are consistent with the hypothesis that the source detected at the location of candidate ``\textit{c}" being a scattered-light disk artifact enhanced relative to the disk knot on the opposite side of the major axis by the merger of the Eastern inner disk rim with the northeastern spiral arm. Aggressive data processing appears to be the main culprit making this disk feature appear point-like in some reductions.  

\begin{figure}[ht]
\centering
\includegraphics[width=0.5\textwidth]{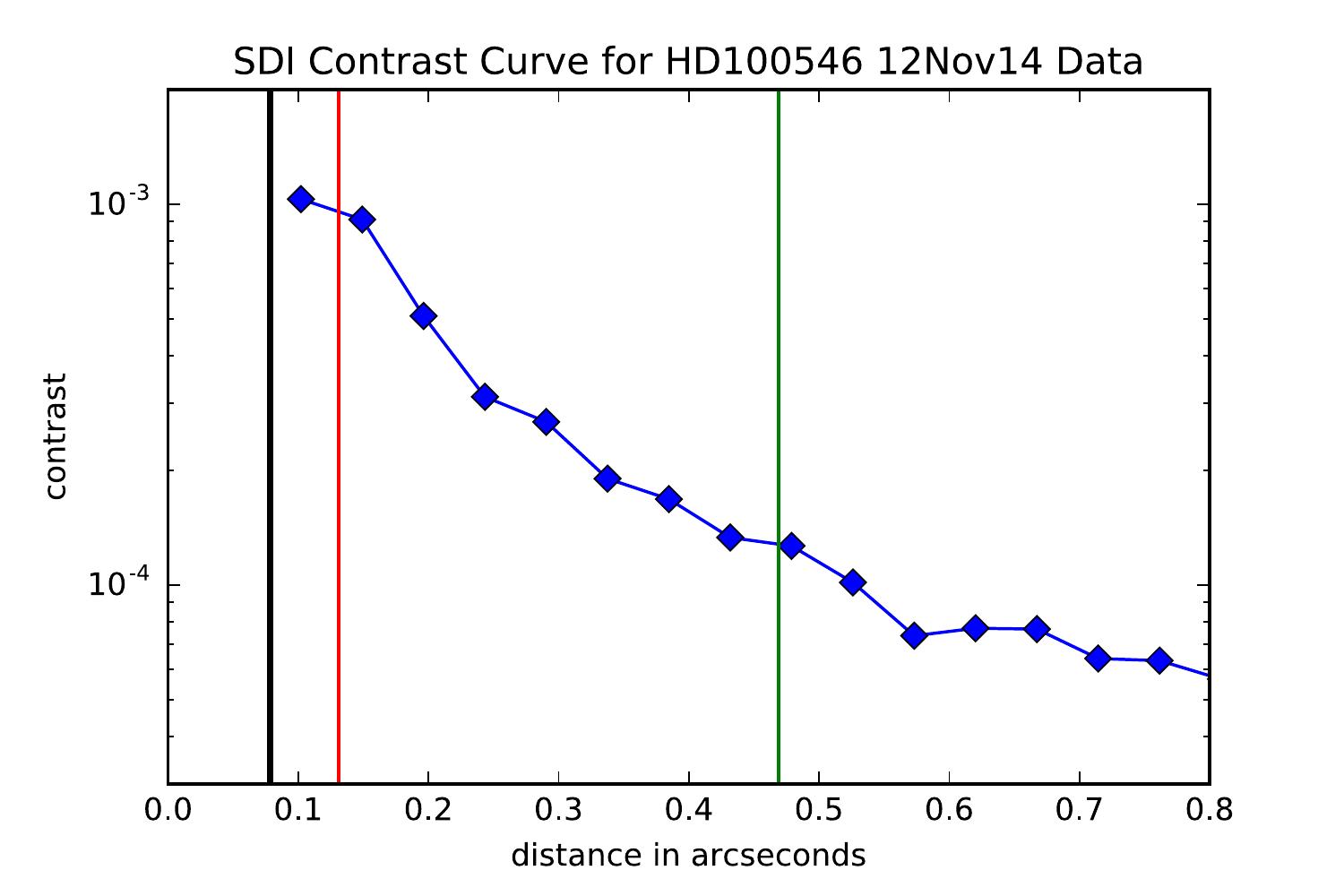}
\caption{Contrast curve for the 2014-04-12 MagAO H$\alpha$ SDI data based and created as described in detail in the text. The thick black line indicates the inner r=8 pixel saturated region of the PSF. The green and red lines indicate the locations of the \textit{b} and ``\textit{c}" planet candidates, respectively. \label{fig:contrast}}
\end{figure}

As a further test of the detectability of an H$\alpha$ point source in this data, we computed an SDI contrast curve, as shown in Figure \ref{fig:contrast}. To compute this curve, we first convolved the final KLIPed image shown in Figure \ref{fig:magao} by a hard-edged circular aperture with a diameter equivalent to the FWHM of the VisAO optical ghost (6 pixels, 0$\farcs$04). This ghost serves as an estimate of the unsaturated PSF of the central star and therefore the size of an independently-sampled region in the image. The convolved image was divided into annuli with widths equivalent to this measured stellar FWHM. Within each annulus, as many independent apertures as would fit in the annulus without overlapping were placed with a random starting point within the annulus, and the central values in these apertures were recorded. The standard deviation of these central values was taken, multiplied by $1/\sqrt{1+1/n}$ (where n is the number of independent apertures) to account for small sample statistics following \citet{Mawet:2014}, and multiplied by 5 to generate the 5$\sigma$ limit for each annulus. 

This procedure was repeated five hundred times (for five hundred random realizations of aperture placements) for each annulus, and the values averaged together. To translate this 5$\sigma$ noise value into contrast, each value was divided by the stellar peak. As HD 100546A was saturated in this dataset, the stellar peak was estimated from a measurement of the ghost peak. Using Moffat fits to the stellar and ghost peaks in five unsaturated GAPlanetS datasets, the ghost was shown to have an intensity equivalent to 0.42$\pm$0.08\% of the stellar peak, and can be scaled by this amount to estimate the stellar peak. 

Finally, throughput was computed by injecting fake planets into the raw H$\alpha$ line images, subtracting the scaled continuum images, and then processing the SDI images with KLIP and the same parameters as the final SDI image. Throughput at a given location is measured as the ratio of the peak brightness of the recovered false planet to the injected planet. The 5$\sigma$ contrast values were multiplied by this throughput to create the final curve. The curve suggests that we could have detected planets up to $\sim1\times10^{-3}$ contrast at the location of the HD 100546 ``\textit{c}" candidate and $\sim1\times10^{-4}$ contrast at the location of HD 100546 \textit{b}. 

HD100546 \textit{b} is heavily embedded in the disk. \citet{Currie:2015} estimate the \textit{H}-band extinction at the location of the point source candidate to be 3.4 magnitudes, which translates to 22 magnitudes of extinction at \textit{R} (and therefore H$\alpha$) following standard Milky Way extinction laws. This is enough to make any constraints on the accretion luminosity of \textit{b} meaningless, as we discuss in more detail in the companion to this paper. 

The ``\textit{c}" candidate, however, is hypothesized to lie at or near the outer edge of the inner disk rim. If it is heavily embedded in the rim (an unlikely hypothesis given the continuity of disk features at this location), then it suffers from the same problem as \textit{b} in that dusty material extincts very efficiently at H$\alpha$ and quickly makes accretion luminosity estimates for embedded protoplanets moot. If the candidate identified by \citet{Currie:2015} or hypothesized by \citet{Brittain:2014} lies inside the cleared central cavity, however, then the contrast limit at this location can be used to place more meaningful limits on the accretion luminosity and accretion rate of any forming protoplanets, albeit with a number of assumptions as detailed below. 

We begin by assuming that the HD100546 cavity is fully cleared of visible light extincting grains, and indeed the precipitous drop in the \textit{Y}-pol radial profile approaching the coronagraph supports this assumption somewhat. We take the measured \textit{V}-band extinction toward HD100546A ($A_V=0.15$, \citet{Sartori:2003}) and translate it to $A_R=0.11$ magnitudes following standard extinction laws \citep{Cox:2000}. Following \citet{Close:2014}, we use this \textit{R}-band extinction estimate and measured contrast, the zeropoint and width of the H$\alpha$ filter, and the distance to HD100546 to translate the measured contrast to an H$\alpha$ luminosity of $1.85\times10^{-4}L_{\sun}$. If we then assume that empirically-derived $L_{H\alpha}$ to $L_{acc}$ relationships for low mass T-Tauri stars also apply to lower mass objects, then following \citet{Rigliaco:2012}, this translates to an accretion luminosity of 0.41\%$L_{\sun}$. Translation of this quantity to an accretion rate requires assumptions about the mass and radius of the accreting object, and we adopt $1.55R_J$ and 2$M_J$ in this calculation as reasonably representative of the population of planets we might expext to scuplt the disk rim. Then, following \citet{Gullbring:1998}, the accretion luminosity translates to an approximate accretion rate of $\dot M\approx1 \times 10^{-8}M_{\sun}/year$, corresponding to growth of a Jupiter mass planet in 100,000 years. The accretion rate onto the primary star is estimated at $\sim10^{-7} M_{\sun}/year$ \citep{Mendigutia:2015}, placing our limit at $\dot M_{planet}<0.1 \dot M_{star}$. We note that a number of assumptions have gone into this estimate, including that accretion onto protoplanets happens in a steady flow of material and not stochastically, and that it is likely only accurate to within 1--2 orders of magnitude. 

\begin{figure*}
\centering
\includegraphics[width=7in]{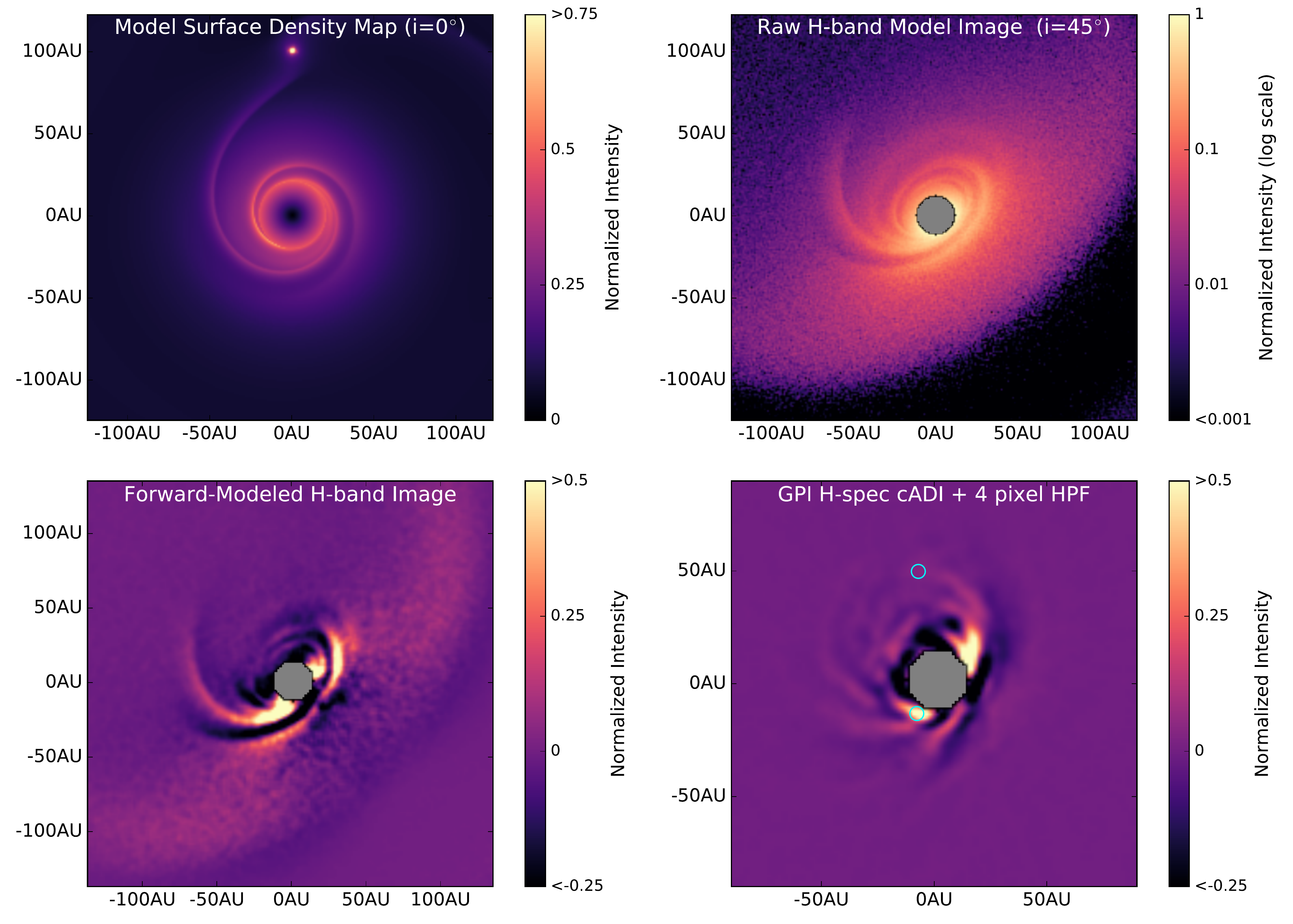}
\caption{Top panel: Density map (left) and Monte Carlo radiative transfer modeled \textit{H}-band image (right) for a planet-induced spiral disk model. The surface density map is shown face-on, and the \textit{H}-band image is for a disk inclined at 45$^{\circ}$ relative to the line of sight and rotated to a major axis PA of 152$^{\circ}$. Bottom panel: Forward modeled \textit{H}-band total intensity image generated by injecting the moded disk into a disk-less GPI dataset with equivalent rotation to the HD 100546 \textit{H}-spec dataset (left) and real HD 100546 data (right). Both were recovered with classical Angular Differential imaging and processed with a 4 pixel high-pass Fourier filter. Although the real disk shows significantly more complex structure than the forward-modeled image, the qualitative similarity is suggestive. \label{fig:model}}
\end{figure*}

\subsection{Disk Modeling}
\label{modeling}

To examine the effects of our data processing procedures on spiral arms, we produce synthetic images of planet-driven spiral arms in disks using combined hydrodynamics and radiative transfer simulations, and process the simulated images using our GPI pipeline. We adopt the 3~$M_{\rm J}$ planet model of \citet{dong16armviewing} with only minor modifications, and briefly summarize salient aspects of the models here. The simulations are described in detail in \citet{dong16armviewing} \citep[see also][]{fung15, dong15spiralarms, dong16hd100453}. The simulations are of spiral arms driven by an outer planetary perturber and do not include an inner disk cavity, though we note the qualitative similarity of spiral arms driven by inner and outer planets demonstrated in other work \citep{Zhu:2015}. We note that the disk models were adopted without modification, and the location of the planetary perturber does not coincide the location of the HD100546 \textit{b} protoplanet candidate. We leave more precise reproduction of HD 100546's specific disk features, including the inner cavity and prediction of the location of planetary perturbers, for future work. 

The three-dimensional density structure of spiral arms in a disk excited by a 3~$M_{\rm J}$ planet was calculated using the code \texttt{PEnGUIn} \citep{fung15thesis}. The initial condition of the disk is $\Sigma\propto 1/r$, and $h/r\propto r^{0.25}$, where $\Sigma$ and $h/r$ are the surface density and aspect ratio in the disk, and $h/r$ at the location of the planet is set to 0.15. The viscosity in the simulation is parametrized using the \citet{Shakura:1973} $\alpha$ prescription with $\alpha=0.01$. The simulation is run for 50 orbits, not long enough for the gap to be fully opened, but sufficiently long for the spiral arms to reach steady state. The resulting 3-D disk density structure is subsequently fed into a Monte Carlo radiative transfer code \citep{whitney13} to produce synthetic $H$-band total intensity images at various inclinations. We convert the gas density as calculated in the hydro simulation to dust density used in the radiative transfer simulation assuming the dust and the gas are well mixed, and we adopt the interstellar medium dust model \citep{kim94} for the dust. These dust grains are sub-um in size, as assumed in previous scattered light spiral arm modeling works (e.g., MWC 758, \citealt{dong15spiralarms}; HD 100453, \citealt{dong16hd100453}).

Planet-induced spiral arms are very robust in scattered light imaging independent of the grain properties assumed in the modeling. Qualitative comparisons such as we are making here are not sensitive to grain models as long as there is small ($\sim\mu$m-sized) dust present in the disk, as modeling of HD 100546's SED suggests is the case (e.g., \citet{Tatulli:2011}).  Additionally, since small grains dominate the opacity at visible and NIR wavelengths and make up the majority of the dust grains in the surface layers of the disk where scattering originates, the assumption of ISM-like dust properties is reasonable. 

To understand the impact of the data processing and qualitatively assess the reality of features identified around HD 100546, the \textit{H} band disk model was convolved with a GPI \textit{H}-band PSF, injected into a GPI datacube of a disk-free star with comparable brightness and a similar amount of on-sky rotation, and then processed via cADI in precisely the same way as the HD100546 GPI \textit{H}-spec data to create a forward model. The underlying surface density model, an \textit{H}-band total intensity model generated via Monte Carlo radiative transfer modeling as described above, and the forward-modeled image are shown alongside the actual on-sky HD100546 cADI \textit{H}-spec image in Figure \ref{fig:model}. 

The forward modeled image suggests that a two-armed spiral disk perturbed by a single planetary companion and viewed at moderate inclination can result in observed structures that are similar in location, number, brightness and extent to the features that we observe in HD100546. This experiment serves as a first-order, albeit striking, demonstration of similarity, and we leave more precise matching and derivation of disk and planet properties from forward models for future work. 

The disk models also naturally produce a near side bright lane feature offset from the rest of the disk and similar in morphology to the feature labeled ``A" in Figure \ref{fig:schematic}. Physically, it corresponds to the outer edge of the bottom side (opposite the disk midplane relative to the rest of the disk emission on both near and far sides) of the disk, and the dark region between it and the other disk features corresponds to the dense disk midplane. This bright lane feature in the raw model and forward-modeled images is beyond the edge of the image in Figure \ref{fig:model}, but it can be seen clearly in Figure 8 of \citet{dong16armviewing}. Tunable model parameters like the thickness of the disk midplane and the scale of the spiral arms could conceivably bring the top side features and the bottom side bright lane feature closer together in modeled images, as they appear to be in HD100546, but we leave this for future work.  Alternatively, bright lane ``A" may correspond to a different variety of disk feature altogether.

Both the forward-modeled and observed images show multiple spiral features, the majority of which lie on the back-scattering far side (NE) of the disk. Self subtraction is clearly seen breaking single spirals from the raw model image into multiple arcs in the forward-model, suggesting that several of the features we identified in Figure \ref{fig:schematic} may belong to contiguous structures. Thus, the forward model also serves to demonstrate the tendency of aggressive PSF-subtraction techniques to create apparent disk clumps along extended features that are smooth in reality, something that will be very important to account for in future studies of planets embedded in circumstellar disks. 

\section{Conclusion}
\label{sec:conclude}

We have presented three new high-contrast imaging datasets for the transitional disk of HD 100546. GPI \textit{Y}-band polarimetric imagery reveals a symmetric disk rim that peaks at 15 au and a spiral arm extending from the Eastern disk rim to the North.  MagAO Simultaneous Differential Imaging at H$\alpha$ (656 nm) and in the neighboring continuum (642 nm) reveal the disk rim, northeastern spiral arm seen in the \textit{Y}-band imagery, and a southern spiral arm that is also present in GPI \textit{H}-band data. 

Deeper GPI \textit{H}-band spectroscopic data allow us to probe outer disk structures, and reveal a number of spiral features in the outer disk. Several outer spiral arms are present in the GPI \textit{H}-band data and, though not revealed in the shallower \textit{Y}-band and MagAO imagery, are similar to structures revealed previously with other high-contrast imaging instruments. These data represent a significant improvement over prior GPI \textit{H}-spec data presented in \citet{Currie:2015} in that they have twice the field rotation and integration time (51.6$^{\circ}$ and 120 min versus 24$^{\circ}$ and 55 min). We find that a large rotational lever arm is extremely important in reliable extraction of the extended features in this very complex disk. 

The lack of planet-like features at the location of HD 100546 ``c" in both H$\alpha$ SDI imaging and in the \textit{H}-band spectra of this region suggest that the apparent point source at this location is an artifact of aggressive processing. This is further supported by the sensitivity of this apparent point source to PSF-subtraction techniques and algorithmic parameters, as well as its location at the intersection between the disk's inner rim and the northeastern spiral arm, where there is a natural concentration of light. 

Finally, we find that the spiral features seen in the disk bear striking similarity to forward-modeled images of a two-armed planet-induced spiral disk at similar inclination. Though we leave detailed extraction of disk and planet properties based on model comparison for future work, we note that the forward-modeled image suggests that the majority of features we've identified are likely real, and several may be pieces of contiguous spiral arms that are separated artificially by disk self-subtraction.  

We believe that this study comprises a cautionary tale, not a prohibitive one. While we have demonstrated that aggressive processing can transform extended disk structures into spurious point-source-like structures, we have also shown that these effects can be mitigated by maximizing field rotation, thoroughly exploring algorithmic parameters, applying multiple PSF subtraction techniques to the same dataset, and comparing structures seen at different wavelengths and with different instruments. As it does not require PSF subtraction, polarized intensity imaging is ultimately the best arbiter of disk morphology. However, lower surface brightnesses in polarized light, the utility of polarized to total intensity comparisons, and the lack of detection of polarized emission from known point-sources suggest that the complete picture of a disk cannot be gleaned from polarized intensity imaging alone. Total intensity disk imaging, as well as the use of aggressive algorithms for PSF removal, will be a continued necessity for the foreseeable future. This study serves to demonstrate that, even with complex and moderately-inclined disks, complementary datasets, thorough exploration of algorithmic approaches and parameters, and deeper observations with maximal field rotation can allow observers to reliably extract high-fidelity disk structures.

\acknowledgements{Based on observations obtained at the Gemini Observatory, which is operated by the Association of Universities for Research in Astronomy, Inc., under a cooperative agreement with the NSF on behalf of the Gemini partnership: the National Science Foundation (United States), the National Research Council (Canada), CONICYT (Chile), Ministerio de Ciencia, Tecnolog\'{i}a e Innovaci\'{o}n Productiva (Argentina), and Minist\'{e}rio da Ci\^{e}ncia, Tecnologia e Inova\c{c}\~{a}o (Brazil). KBF and JF's work was performed in part under contract with the California Institute of Technology (Caltech)/Jet Propulsion Laboratory (JPL) funded by NASA through the Sagan Fellowship Program executed by the NASA Exoplanet Science Institute. KBF and BM's work was supported by NSF AST-1411868. Portions of this work were performed under the auspices of the U.S. Department of Energy by Lawrence Livermore National Laboratory under Contract DE-AC52-07NA27344. KMMmax, TB, and LMC's work is supported by the NASA Exoplanets Research Program (XRP) by cooperative agreement NNX16AD44G. JRG, RDR, PK, JW, VB and other members of the GPIES team are supported by NASA grant number NNX15AD95G. Support for MMB's work was provided by NASA through Hubble Fellowship grant \#51378.01-A awarded by the Space Telescope Science Institute, which is operated by the Association of Universities for Research in Astronomy, Inc., for NASA, under contract NAS5-26555}

\facility{Gemini: South (GPI) and Magellan: Clay (MagAO)}

%\bibliography{ref.bib}

%% This command is needed to show the entire \author+affilation list when
%% the collaboration and \author truncation commands are used.  It has to
%\all\authors


\end{document}